\documentclass[aps,pre,groupedaddress,showpacs]{revtex4-1}
\usepackage{amsmath,bm}

\usepackage{hyperref}
\usepackage{color}

\newcommand{\be}{\begin{equation}}
\newcommand{\ee}{\end{equation}}
\newcommand{\bea}{\begin{eqnarray}}
\newcommand{\eea}{\end{eqnarray}}

\DeclareMathOperator{\sign}{sign}

\usepackage{tikz}
\usetikzlibrary{arrows,chains,matrix,positioning,scopes}
\makeatletter
\tikzset{join/.code=\tikzset{after node path={%
\ifx\tikzchainprevious\pgfutil@empty\else(\tikzchainprevious)%
edge[every join]#1(\tikzchaincurrent)\fi}}}
\makeatother
\tikzset{>=stealth',every on chain/.append style={join},every join/.style={->}}
\tikzstyle{labeled}=[execute at begin node=$\scriptstyle,execute at end node=$]

\begin{document}
\author{Andrea Baldassarri}
\title{Universal Excursion and Bridge shapes in ABBM/CIR/Bessel processes}
\affiliation{Istituto dei Sistemi Complessi - CNR and Dipartimento di Fisica, Universit\`a di Roma Sapienza, P.le Aldo Moro 2, 00185, Rome, Italy}

\begin{abstract}
{Several years ago, in the context of the physics of hysteresis in magnetic materials, a simple stochastic model has been introduced: the ABBM model. Later, the ABBM model has been advocated as a paradigm for the description of a broad class of diverse phenomena, baptized "crackling noise phenomena". The model reproduces many statistical features of such intermittent signals, as for instance the statistics of burst (or avalanche) durations and sizes, in particular the power law exponents that would characterize the dynamics as critical. In order to go beyond such "critical exponents", the measure of the average shape of the avalanche has also been proposed.
Here, the} exact calculation of the average (as well as the fluctuations) of the avalanche shape for the ABBM model is presented, showing that its normalised shape does not depend on the external drive. Moreover, the average (and the fluctuations) of the multi-avalanche shape, that is the average shape of a sequence of avalanches of fixed total duration, is also computed. Surprisingly, the two quantities (avalanche and multi-avalanche normalised shapes) are exactly the same. This result is obtained using the exact solution of the ABBM model, which is obtained leveraging the equivalence with the Cox-Ingersoll-Ross process (CIR), rigorously obtained with a so called "time change". A simple presentation of this and other known relevant exact results is provided: notably the correspondence of the ABBM/CIR model with the Rayleigh model and, more importantly, with the generalised Bessel process, which describes the dynamics of the modulus of the  multi dimensional Ornstein-Uhlenbeck process (exactly as the Bessel process does for the Brownian process). As a consequence, our main finding, that is the correspondence between the excursion (avalanche) and bridge (multi-avalanche) shape distributions, turns to apply to all the aforementioned stochastic processes. In simple words: if we consider the distance from the origin of such diffusive particles, the (normalised) average shape of its trajectory (and the fluctuations around that) until a return in a time $T$ is the same, whether it has returned before $T$ or not.
\end{abstract}

\maketitle

\tableofcontents

\section{Introduction}

Stochastic processes are a fundamental tool in physics, a notable example is the celebrated Einstein work on Brownian motion, which anticipated the development of stochastic calculus and the theory of continuous Markov processes. Nowadays, we witness a surge of interest for such processes in physics, in order to better tackle and ground the study of non-equilibrium phenomena~\cite{Seifert2012,VanDenBroeck2012}.

A succesfull example of a simple stochastic model was introduced many years ago to describe the phenomenon of Barkhausen effect (see Fig.~\ref{hysteresis}), that is the the very irregular noise generated by ferromagnetic material under a varying applied field (see{~\cite{Bertotti2005}, and also}~\cite{Colaiori2008a} for a theoretically oriented review on the subject). In order to describe the observed phenomenology and give a physical rationale, Alessandro, Beatrice, Bertotti, and Montorsi proposed a stochastic phenomenologic model,  the ABBM model~\cite{Alessandro1990}, by the initial of its authors, which has later been advocated as a successful effective one dimensional version~\cite{Zapperi1997,Zapperi1998} of a more general class of models for "crackling noise"~\cite{Sethna2001}, that is the appearance of intermittent, bursting temporal measurements in a bunch of diverse natural phenomena, ranging from dislocation dynamics to earthquakes.  The model is often presented as a "mean-field" model for the problem of dynamics of an elastic interface in a random pinning potential~\cite{LeDoussal2012}, 
which would justify the apparent universality of the model phenomenology (see ~\cite{Wiese2021} for a recent review on theory and experiments of elastic manifolds, with an extensive bibliography, and ~\cite{TerBurg2020} for a criticism of the mean-field character of the ABBM model).

Generally speaking, a "crakling-noise" dynamics can be described as a sequence of avalanches of activity, whose statistics can be characterized by several probability distributions (as for instance of durations or sizes) which present a regime of  algebraic decay, identifying some "critical", hopefully universal, exponents~\cite{Bohn2018}. 

\begin{figure}[ht]
  \includegraphics[width=12.0cm]{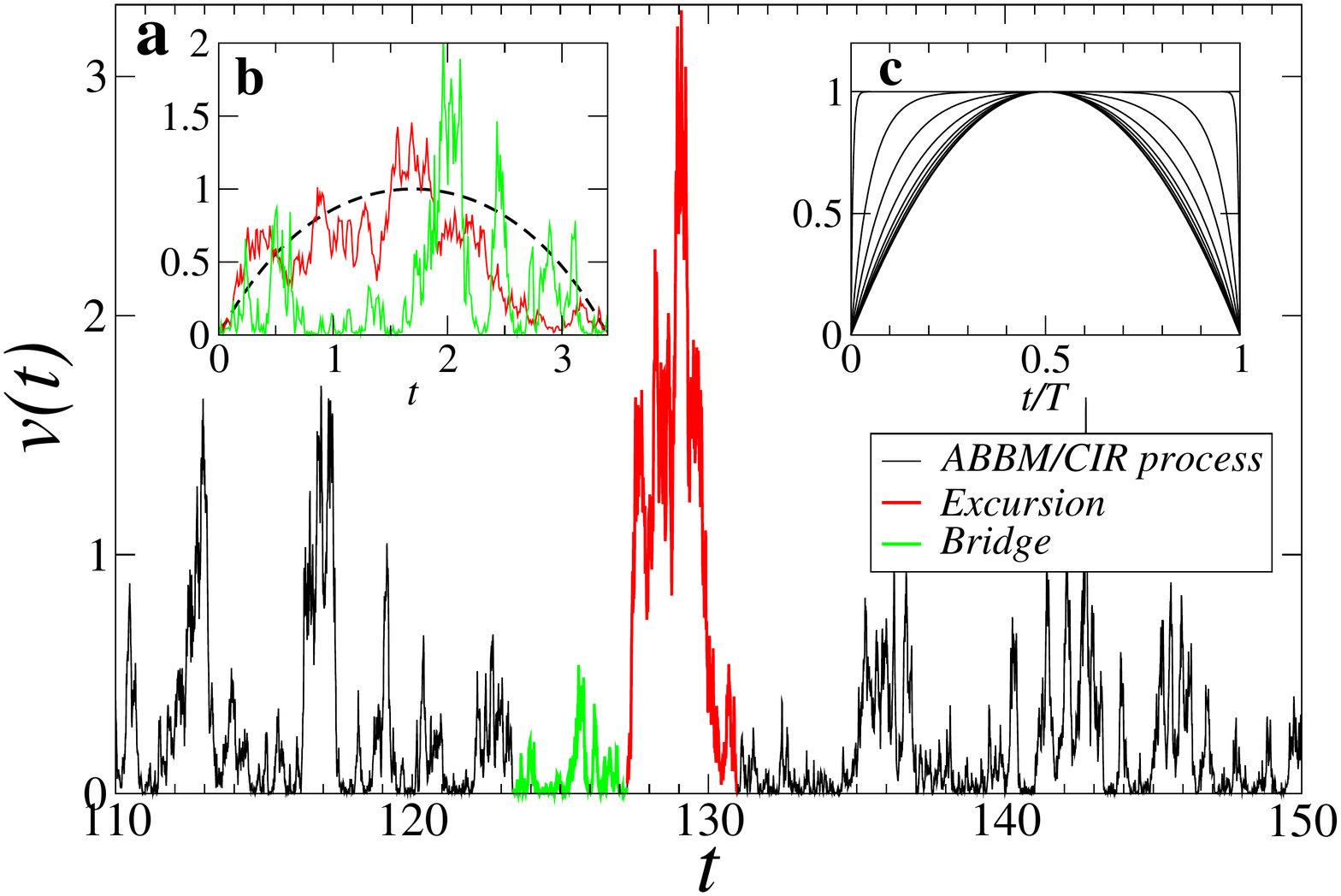}
  \caption{Pictorial representation of the subject of the paper. (a) Single realization of a trajectory of the ABBM/CIR process. In red an avalanche (excursion) of the process, in green a multi-avalanche (bridge) of comparable length. (b) Excursion and bridge of plot (a) are rescaled to compare with their respective normalised average. (c) The normalised average shape of the excursion and bridge, for different durations $T$ as a function of the rescaled time $t/T$ (see Eq.~\ref{CIRavex}). The shape changes, from a parabolic profile for small durations $T\ll 1/k$, to a flat profile for very large durations $T\gg 1/k$. \label{fig1}  }
\end{figure}

In order to further inspect the dynamics, other quantities have been investigated, beyond exponents. This is the case of the average avalanche shape. The idea is to consider the set of avalanches of similar durations, and to average their profile, that is the value of the signal at a fixed time $t$ after the beginning of the avalanche  (see Fig.~\ref{fig1}).
In the case of the Barkhausen noise, this quantity has been measured in several papers~\cite{Spasojevic1996,Kuntz2000,Mehta2002,Durin2002} and then theoretically computed, with several degrees of approximations, in a number of papers~\cite{Baldassarri2003,Colaiori2004,Colaiori2008a,Papanikolaou2011,LeDoussal2012}. 

Average avalanche shapes have also been investigated in bursting signals from a variety of
materials, well beyond magnetic systems~\cite{Papanikolaou2011,Bohn2018}, ranging from
intermetallic compounds and crystals~\cite{Chrzan1994,Sparks2018,Sparks2019} to glassy and amorphous systems~\cite{antonaglia14,ferrero16,Lagogianni2018,laurson13}, granular materials~\cite{Baldassarri2019}, quasi brittle materials~\cite{Vu2020},  and, very recently,  in cortical bursts~\cite{Roberts2014,Wikstro2015}, in transport processes in living cells~\cite{Danku2013}, as well  in
ants~\cite{Gallotti2018} and  in  human~\cite{Chialvo2015} activity.  
Burst shape  has also been investigated  in stellar processes~\cite{Sheikh2016}, Earth's
magnetospheric dynamics~\cite{Consolini2008}, earthquakes~\cite{Metha2006}

Here, we give an exact and complete computation of the average avalanche shape (known as average "excursion" in the theory of stochastic processes), as well as its fluctuations, for the ABBM model.  The computation is extended to the statistics of the "multi-avalanche" (or "bridge") shape, that is the train of avalanches of a fixed total duration (see Fig.~\ref{fig1}). This quantity, which has never been considered before for the ABBM model, could be more suitable to be measured in a stochastic signal, providing a better statistics, especially for large durations. The two quantities (avalanche and multi-avalanche shape) reveal a quite unexpected universality in their normalised analytical form. 

The exact computations are performed leveraging  the equivalence of the ABBM model with the Cox-Ingersoll-Ross model (CIR)~\cite{Cox1985}, a standard stochastic process in finance~\cite{Jeanblanc2009}. Such equivalence, which has been informally derived before~\cite{Papanikolaou2011}, can be rigorously proven using the theory of {\em random time change}~\cite{Bjork2019}, which is briefly introduced below. Exploiting this technique, we summarise the main relations between the ABBM/CIR model and other stochastic process, as the Rayleigh process, and more generally with the generalized Bessel processes~\cite{Going-Jaeschke2003}, which in turn can describe the statistics of the modulus of a multi-dimensional Ornstein-Uhlenbeck process.

In order to have an intuitive rationale of the equivalence between avalanche (excursion) and multi-avalanche (bridge), we consider the stochastic equations for the respective constrained process via the so called Doob's h-transform. In this framework, the excursion differs from the bridge because of the presence of a repulsive drift from the origin, which appears to be irrelevant to the statistics of the normalised shape. For a strictly related class of process, the Bessel process and the multi-dimensional radial Ornstein-Uhlenbeck process, the additional term of the excursion stochastic equation  is not a constant drift, but rather a "logarithmic potential force". Nevertheless, the normalised shape of the excursion is again identical to the average bridge normalised shape.

In conclusion, the equivalence of the normalised shapes of bridge and excursion applies to the whole class of diffusing processes mentioned above. For specific values of the parameters, they represent the dynamics of the distance from the origin of a diffusing particle in a quadratic well (in the over-damped limit). In this context, the results presented here prove a quite unexpected feature: the (normalised) average shape of the radial trajectory up to a return to the origin, does not depend on the previous number of returns to the origin.

The paper is organised in the following way. In Section~\ref{ABBMsection} we introduce the ABBM model and we sketch the connection with the CIR stochastic equation. In Section~\ref{TimeChangeSection}, we briefly introduce the random time change. In Section~\ref{CIRSection}, the random time change technique is used to connect the ABBM model with other famous stochastic processes (see Table~\ref{FigStochasticProcesses}). In Section~\ref{CIRSolutionSection} we review the probabilistic relevant solutions for the CIR model and discuss their main properties. In Section~\ref{ExcursionBridgeSection} we exactly compute the statistics of durations and shape of excursions and bridges, and show the universality in the normalised shapes. This section contains the main original and novel results of the work. In section~\ref{DoobsSec} we consider the stochastic differential equation for the excursion and the bridge of the CIR process, and we recover the average shapes computed before. Furthermore, we extend the computation to the Bessel process and its squared, and show that the universality holds even in this case. Finally, a Conclusion Section ~\ref{ConclusionsSec} summarizes and discusses the results.
 

\section{ABBM Model} \label{ABBMsection}

{ Applying a magnetic field to a ferromagnetic material under their Curie temperature, hysteresis phenomena occur: the response of the material (its magnetization) depends on the hystory of the applied field and a graph of magnetization versus a periodic field shows a wide range of curves known as hysteresis loops. This observation is at the very basis of modern theory of phase transitions. However the actual phenomenology of hysteresis comprehend dynamical phenomena which need a more detailed description of the material with respect to a pure Ising model. In particular, the variation of the magnetization under a slowly increasing or decreasing applied magnetic field may not be smooth, but rather proceed with irregular jumps or a sequence of small abrupt variations. In some cases, this irregular response of the materials can even be amplified to be appreciated as an acoustic, crackling noise, as was discovered by Barkhausen using the speaker of a telephone (an english translation of the original Barkhausen's paper can be found as an appendix in~\cite{Durin2005}). See Fig.~\ref{hysteresis} for a graphical illustration of the Barkausen effect.

\begin{figure}[ht] \centerline{\includegraphics[width=5.0cm]{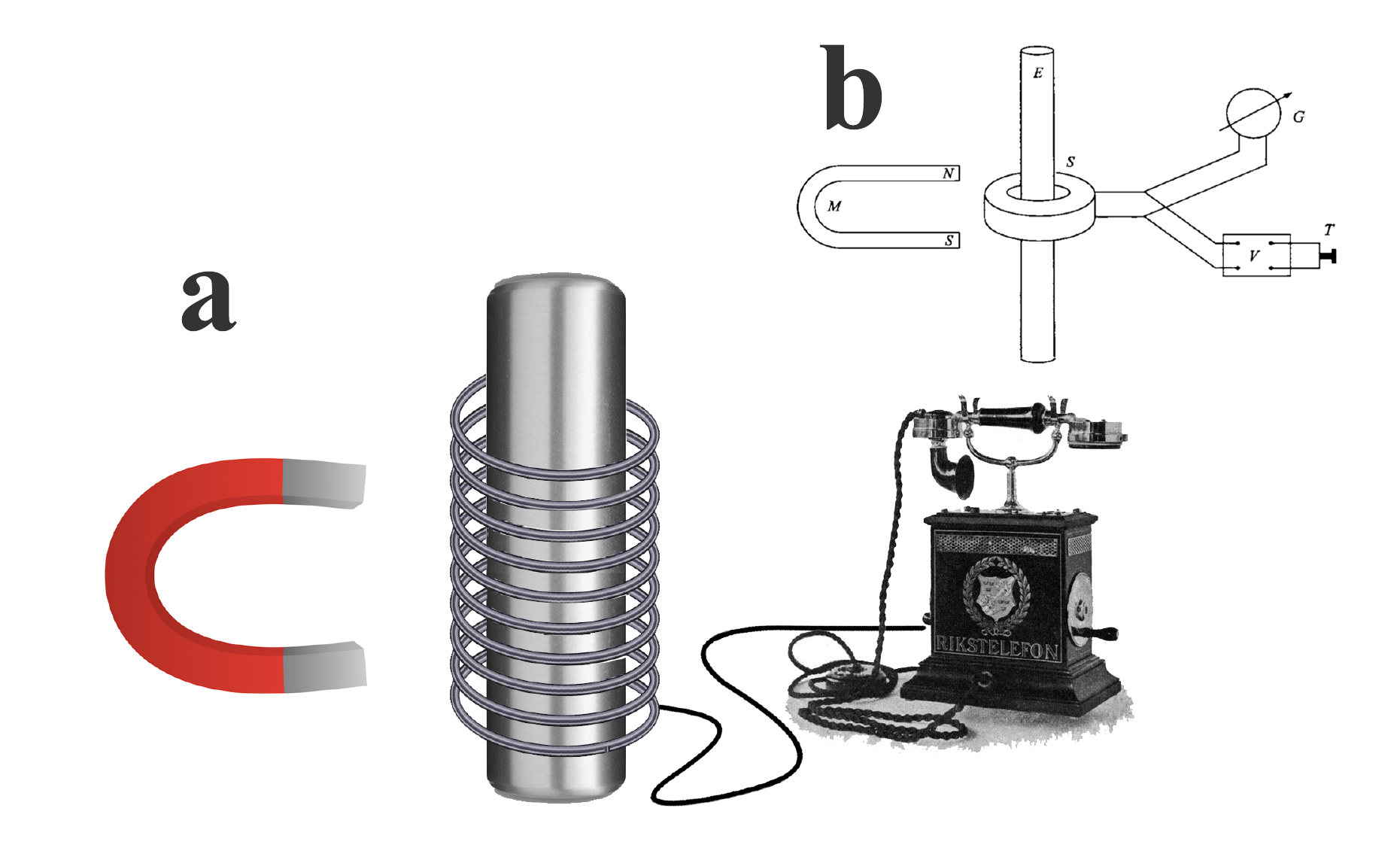}
  \includegraphics[width=10.0cm]{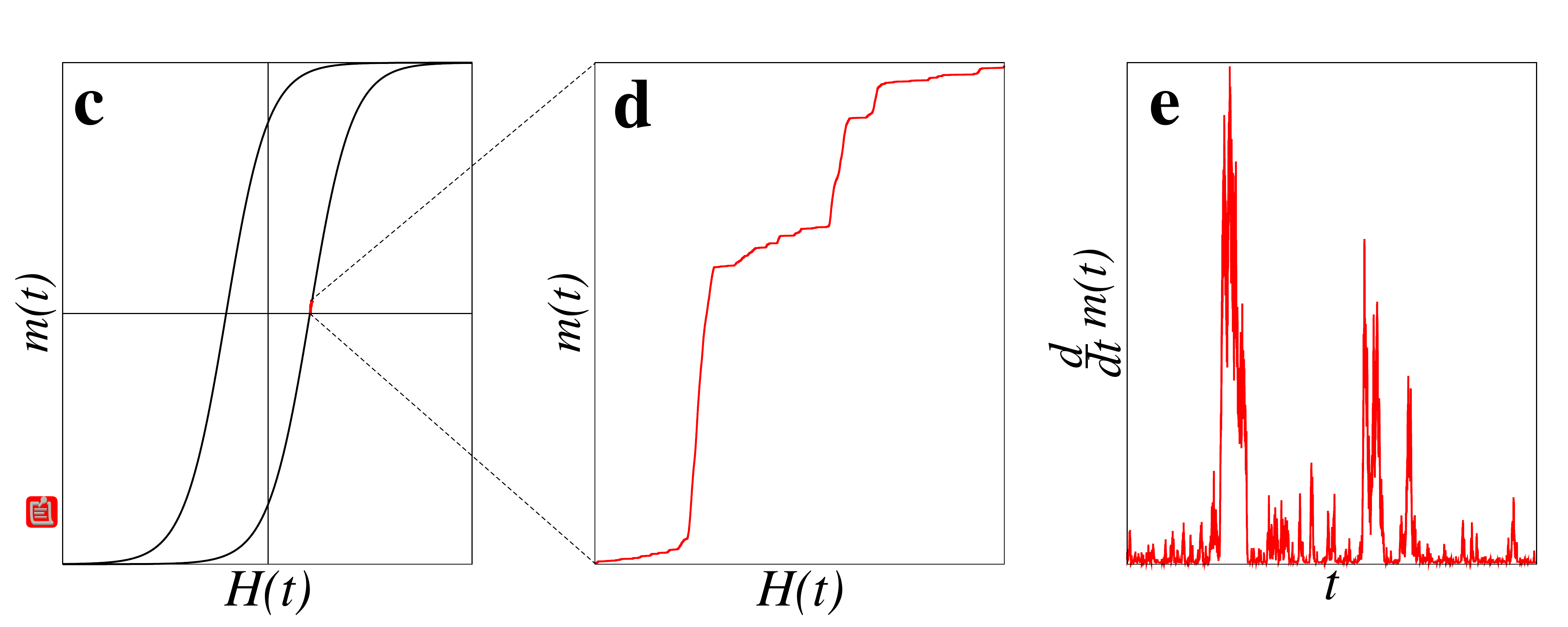}}
  \caption{{Pictorial representation of: the original Barkhausen experiment (a), with the diagram appeared in his original paper (b), and an ideal hysteresis loop (c), with a zoom displaying the Barkhausen effect (d), and the Barkhausen noise (e). \label{hysteresis}}}
\end{figure}

The  ultimate reason for such irregular, sample dependant response, relies in the random impurities present in the materials, ascribing the phenomenon in the physics of disordered systems.} The first comprehensive statistical theoretical approach to Barkhausen noise was done in 1990 by Alessandro, Beatrice, Bertotti and Montorsi, who proposed a phenomenological model, which was then named after the authors of the two companion papers, the first of which dealt with theory~\cite{Alessandro1990}, the other with experiments~\cite{Alessandro1990a}.  
The ABBM model was inspired by the work of N\'eel~\cite{Neel1954}, who was the first to introduce a random energy model into the study of hysteresis. In the Alessandro et al. approach, the idea of a random energy landscape is generalized, and used to construct a stochastic equation for the domain wall dynamics, in order to describe the Barkhausen effect.

\begin{figure}[ht] 
  \includegraphics[width=12.0cm]{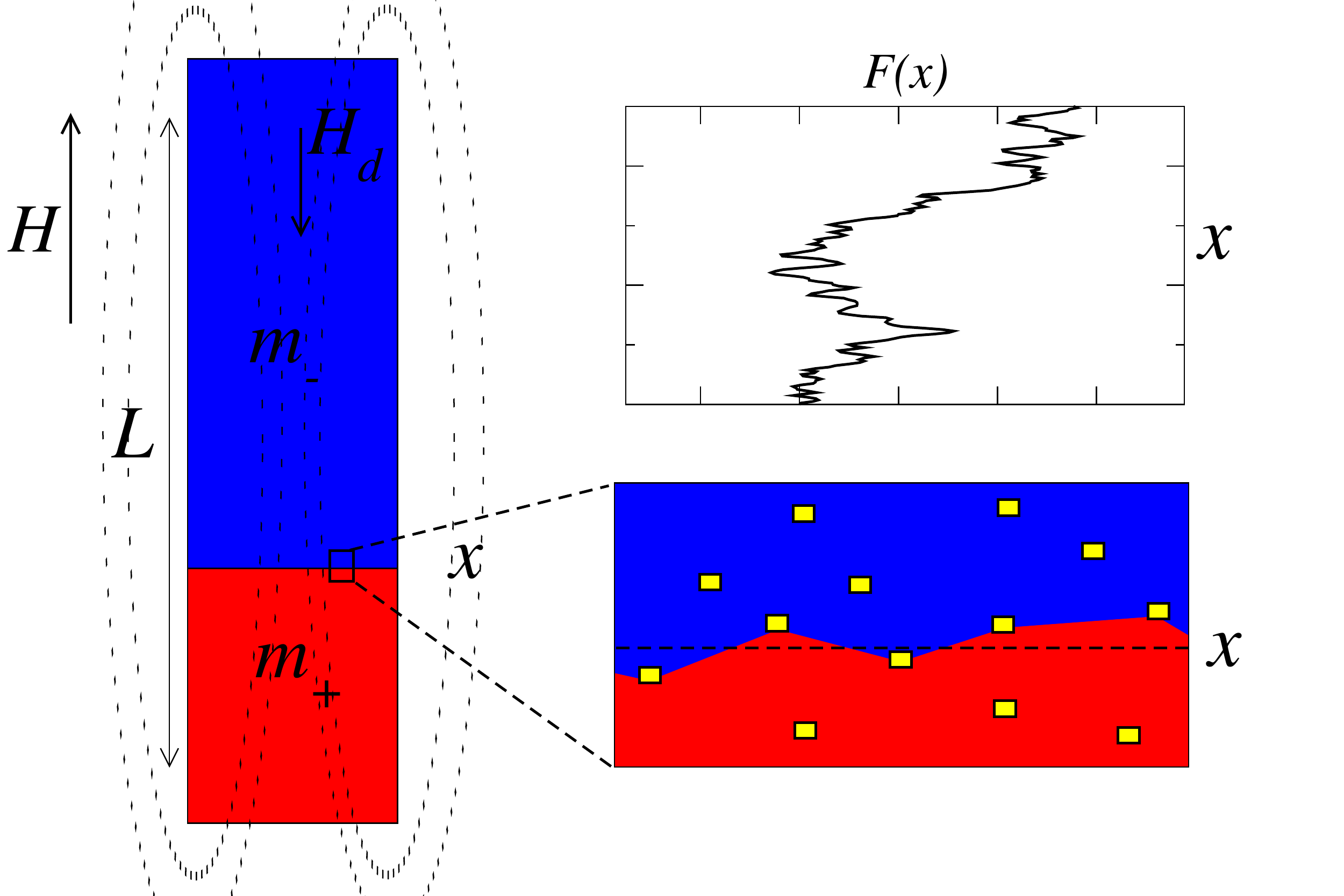}
  \caption{Phenomenolgical idealization of a single domain wall magnet\label{domainwall}  }
\end{figure}

{ The ABBM model is a phenomenological model based on a simplified description of the magnetic response of the sample subject to an applied external field $H$.  Note that the effective field inside the sample depends on the magnetostatic response of the sample, and its actual value should be evaluated with an explicit solution of the macroscopic Maxwell equations, that is considering the magnetizing field inside the material, taking care of the conditions at the border of the sample.  For instance, in a case of an ideal (i.e. without impurities) horseshoe shaped sample the effective magnetic field acting in the material is $H - H_d$, where $H_d$ is the so called demagnetizing field, that in this case can be exactly computed and is $H_d = N m$, where $m$ is the sample magnetization and $N$ is a constant depending on the sample sizes. The computation of the demagnetezing field for a generic geometry is a non trivial task. However, the ABBM model assumes that the demagnetizing field is proportional to the sample magnetization, exactly as in the case of the horseshoe geometry, which is in general a quite satisfactory approximation for usual experimental setups~\cite{Bertotti2005}.  

Moreover, ABBM model assumes the presence of two competing magnetic domains inside the material characterized by a magnetization density $m_+$ and $m_-$, that is it assumes the existence of a single planar, transversal domain wall, as depicted in the sketch of Fig.~\ref{domainwall}. The total magnetization of the sample depends on the average position $x$ of the wall, as
\be
m(x) =  V \left[\frac xL m_+ + \frac{L-x}L m_-\right],\label{SingleDomain}
\ee
where V is the sample volume. 

The domain wall is subjected to the effective field considered above, but it also also experience the effect the impurities inside the material that contribute as a random magnetic pinning field. Since the impurities are a quenched randomness inside the sample, i.e. does not change in time, the random field depends on the position of the domain wall $x$. The statistical characterization of the random field $H_p(x)$ is a crucial ingredient of the model and it will be discussed below.

The motion of the domain wall, is described by a single scalar velocity $v = \frac{dx}{dt}$, which is proportional to the Barkhausen signal $\frac{dm}{dt}$ (see Fig.~\ref{hysteresis}, panel e):
\[
\frac{dm(t)}{dt} = \frac VL (m_+ - m_-) \, v(t)
\]
The wall velocity, in turn, is directly related to the total magnetic forces acting on it, according to a overdamped approximation of its dynamical equation, which disregard inertial or memory effects:
\begin{equation}
v(t) \propto H - H_d + H_p(x).
\end{equation}

Considering a constant applied field $H_0$, the domain will reach an equilibrium position $x_0$, such that the net force is zero: $H - N m(x_0) +H_p(x_0)=0$.
If we now start to slowly increase the applied field, at a constant rate $\frac{dH(t)}{dt}  = h $, we obtain 
\begin{equation}
v(t) \propto H_0 + h t  - \frac{N(m_+-m_-)}{V}L x - \frac{N}V m_- + H_p(x). \
\end{equation}

Eploiting the initial condition $H_0 = m(x_0)-H_p(x_0)$, one gets
\begin{equation}
v(t) \propto h t  - \frac{N(m_+-m_-)}{VL} (x-x_0) + H_p(x)-H_p(x_0). \label{overdampedDW}
\end{equation}
If, without any loss of generality, we change the notation $x-x_0\to x$ and $H_p(x)-H_p(x_0)\to F(x)$, 
the overall result of the previous assumptions is that Eq.~\ref{overdampedDW} is now a stochastic differential equation for the average position of the domain wall, that can be written as
\begin{equation}
\frac{dx}{dt} = k\left[c t - x\right] + F(x)\label{ABBM}.
\end{equation}
where $k$, and $c$ are constant parameters of the equation, which depends on the experiment specific quantities ($N$, $V$, $m_+$, $m_-$, $h$, $V$, $L$, etc...).

This equation resembles a Langevin equation for an overdamped Brownian particle submitted to a time varying drift. However, the usual thermal noise term, the source of stochasticity, is now replaced by a random force $F$ which depends on the position $x$, instead of the time $t$. This is due to the fact that the impurities represent a form of quenched disorder and do not change in time. The stochasticity of the equation comes from the fact that the system is driven by the increasing external field, which moves the domain wall, and at each new position $x$ it experiences a new random value of the force $F(x)$, which depends on the contribution of the impurities that interact with the domain wall given its position $x$. 

In average the contribution of such force is zero:
\begin{equation}
\overline{F(x)} =0,\label{BF1}
\end{equation}
where the average is over the possible position of the domain wall $x$. This is the first statistical characterization of the random force, which simply corresponds to an homogeneous densities of impurities inside the material. 

More crucial is the characterization of the spatial correlations of the random force. 
Inspired by some experimental investigation on systems with a single domain wall~\cite{Baldwin1972,Grosse-Nobis1977,Vergne1981}, the ABBM model assumes that the spatial correlations of the force is:
\begin{equation}
\overline{ \left[F (x) - F (x+y)\right]^2 } = \sigma^2 |y|\label{BF2}
\end{equation}
In other words, $\frac{dF}{dx} = \eta(x)$ is a white noise (in spatial coordinate): $\langle \eta \rangle = 0$ and $\langle \eta(x)\eta(x+y)\rangle = \sigma^2 \delta(y)$, or,  in mathematical terms $F$ is a Brownian process (along the $x$ parameter). 

There is an intuitive rationale for such statistical characterization of of the random force $F$.  Since  the random force is the sum of the contribution of many independent impurities, it is reasonable to assume it as Gaussian distributed. Moreover, a small variation of the the average domain wall position $dx$, may be considered as the result of the depinning of the domain wall from the attraction of some, but not all the of the impurities it was interacting with, and the encounter with some new pinning impurities. Being the impurities homogeneously distributed in space, the number of impurities, and consequently the force experienced by the wall, will fluctuate as $\sqrt{dx}$. 

Consistently with such wave hand waving  argument, the ABBM model has been shown to represent the limit of a problem of dynamics of an elastic interface in a random pinning potential~\cite{LeDoussal2012,Wiese2021}. More precisely, consider a driven elastic interface embedded in a $d$ dimensional space. In absence of overhangs, it can be described by a displacement field $u(y,t)$, where $y$ now are $d-1$ space coordinates ortogonal to the driving direction. Its dynamical equation reads, in the overdamped approximation:
\be
\partial_t u(y,t) = (\nabla^2_y - k) (u(y,t)-c t) + f(y,u(y,t)),\label{FRG}
\ee
where $f(y,u)$ is the random force in the point of the $d$-dimensional space of coordinates $(y,u)$, due to the presence of pinning disorder.  $k$ is a measure of a restoring force which flatten the interface beyond a scale $1/\sqrt{k}$ and $c$ is the average velocity of the interface, and it is proportional to the constant increasing rate of the driving force.

This is a very general model for the depinning transition of an elastic manifold in a disordered medium and Eq.~\ref{ABBM} appears an effective, one dimensional equation for the "center of mass" of the interface
\[
x(t) = L^{1-d} \int u(y,t) dy
\]
($L^{d-1}$ is the extension of the interface). $F(x)$ is the effective random force experienced by interface when its center of mass is at the displacement $x$. It turns out that the statistical characterization of the random force $F$ adopted by the ABBM model, appears very "judicious"~\cite{LeDoussal2012inertial}, at least for small displacements, since it coincides with the tree approximation of a functional renormalization group analysis of the original problem~\eqref{FRG}, and in this sense it is a mean field model for very slow driving $c=0^+$.

As explained above, the stochastic equation defining the ABBM model has no thermal noise. From this point of view it can be seen as a zero temperature limit of a more general random walk in a special random environment, where the force is a Brownian process (at odds for instance with the Sinai model~\cite{Sinai1983}, where the force is derived from a Brownian potential).}

A possible generalization of the model can take inertial effects into account:
\begin{equation}
(I\partial_t v) +  v = k(c t - x) + F(x)\label{inertialABBM}
\end{equation}
where, again, the stochastic force has zero average and "Brownian" spatial correlations. The original ABBM model is recovered in the over-damped limit of small inertia $I\to 0$. An ABBM model with inertia has been studied in~\cite{LeDoussal2012inertial} and previously introduced~\cite{Baldassarri2006} to sucessfully reproduce the statistics of stick-slip dynamics in granular friction experiments~\cite{Dalton2001,Dalton05,Petri2008,Annunziata2016}, for which the average avalanche shape has been also measured and discussed~\cite{Baldassarri2019}.  Different inertial terms have also been considered~\cite{Zapperi2005,Dobrinevski2013}. The case of over-damped dynamics, but with a non-stationary driving has been considered in finance~\cite{Maghsoodi1996,Shirakawa2002} and in physics~\cite{Dobrinevski2012}. Here we'll stick the discussion on the standard over-damped ABBM model, defined by Eqs.~\eqref{ABBM}, ~\eqref{BF1}, and ~\eqref{BF2}.

Before entering in a detailed exact discussion of the stochastic equation, let make some qualitative observation. First let us note that the velocity keeps always non-negative values $v(t)\ge0$ during the dynamics. In fact, being $v(t)$ a continuous function of time, if at $t=t_0$ at a certain position $x(t_0)=x_0$, it happens that $v(t_0)=0$, then at time $t=t_0+\epsilon$, one has  
\[
v(t_0+\epsilon) = k c \epsilon + O(\epsilon^2) > 0
\]
where we used that $F(x_0) = -k(c t_0 - x_0)$.

As we will explain in more rigorous way in the next sections, in this case the equation can be recast~\cite{Papanikolaou2011,Dobrinevski2012} in a more comfortable stochastic differential equation (sde), of the form:
\begin{equation}
dv = k(c-v)dt + \sigma \sqrt{v} \,dW_t \label{CIRsde}
\end{equation}
where $W$ is the usual Wiener process.
Note that such equivalence was claimed in the very first formulation of the ABBM model~\cite{Alessandro1990}, where the sde equation is written as:
\[
\frac{dz}{dt} + \frac{z-c}\tau = \frac{dw}{dt}
\]
and the noise $w(t)$ is characterised by the following informal relation $\langle |dw|^2\rangle = 2z \frac{dt}\tau$. Such a characterization of the  noise term can be rigorously recast through a mathematical procedure known as {\em time change}~\cite{Jeanblanc2009}, for which we give hereafter a intuitive demonstration, which will be useful in the rest of the paper.

\section{Time changes}~\label{TimeChangeSection}

The problem of time change is the following: suppose you have a (continuous Markov) stochastic process defined by the equation (in the It\^o's scheme of calculus):
$$
dX =  a\left(X,t\right) dt+ b\left(X,t\right)dW_t.
$$
Now consider an increasing, continuous, increasing function of time $\tau(t)$. We would like to use $\tau$ as a clock for our stochastic process, instead of $t$.  What is the new stochastic equation? That is what are $\hat a$ and $\hat b$ in
$$
dX = \hat a\left(X,\tau\right) d\tau + \hat b\left(X,\tau\right) dW_\tau
$$
Since in principle $\tau(t)$ can be a stochastic process, we write
\begin{equation}
d\tau = G^2(X,t) dt,\label{timechange}
\end{equation}
where $G^2(X,t) =  \frac{d\tau(t)}{dt} \ge 0$ if $\tau(t)$ is a purely deterministic (increasing) function, while $G(X,t) =  g(X)$ is a smooth, non negative function if we consider a random time change. In both case, the solution is
\begin{equation}
\tau(t) = \int_0^{t} G^2\left(X(s),s\right) ds\label{rtc}
\end{equation}
and can be formally inverted 
$$
t(\tau) = \inf \left\{ s\ge 0 : \int_0^s G^2\left(X(u),u\right) du = \tau\right\}.
$$

With this definitions, the drift term of our new stochastic equation is obviously {$\hat a\left(X,\tau(t)\right) = a\left(X,t\right)/G^2(X,t)$}. For the diffusive term, instead, we need to deal with $dW_{t(\tau)}$ which should now be written in terms of $dW_\tau$.
The self-similarity of the Wiener process,  $d W_{c\,t} = \sqrt{c}\, dW_t$, implies that the solution is the following:
$$
dW_{t(\tau)} = G(X,t)\, dW_{\tau}.
$$

Using this result, we get the solution to our problem:
\begin{eqnarray}
\hat a(X,\tau) &=& \frac{a\left(X,t(\tau)\right)}{G^2(X,t)}\\
\hat b(X,\tau) &=& \frac{b\left(X,t(\tau)\right)}{G(X,t)}
\end{eqnarray}

Furthermore, considering together with the time change Eq.~(\ref{timechange}), a rescaling of the random variable, that is a transformation 
\[
Y(\tau) = f(\tau) X(t(\tau)),
\] it is easy to show (using It\^o's lemma~\cite{Gardiner1985}) that the coefficients for the equation 
$dY = \hat a\left(Y,\tau\right) d\tau + \hat b\left(Y,\tau\right) dW_\tau$
are:
\begin{eqnarray}
\hat a\left(Y,\tau\right)&=& Y \frac{d \ln(f)}{d\tau} + a\left(\frac{Y}{f(\tau)},t(\tau)\right) \frac{f(\tau)}{G^2(X,t(\tau))}\label{tc1}\\
\hat b\left(Y,\tau\right)&=& b\left(\frac{Y}{f(\tau)},t(\tau)\right) \frac{f(\tau)}{G(X,t(\tau))}\label{tc2}
\end{eqnarray}

{To our knowledge, this useful technique has been seldom exploited by the physicist community. It has been recently utilized in the field of stochastic thermodynamics~\cite{Pigolotti2017}, in order to give a unified and refined version of several inequalities for entropy production and other thermodynamic quantities, as the "housekeeping heat" for stationary non equilibrium systems~\cite{Chetrite2018,Neri2019,Chun2019}.}

\subsection{Example: Ornstein-Uhlenbeck}

Consider the (deterministic) time change of the Wiener process, with $f(\tau)=e^{-k\tau}$
and $t(\tau) = \frac{\sigma^2(e^{2k\tau}-1)}{2k}$, that is the process:

\begin{equation}
Y(\tau) = e^{-k\tau} W\left(\frac{\sigma^2 (e^{2k\tau}-1)}{2k}\right).\label{Winer2OUTimeChange}
\end{equation}

Using the above formulas Eq.~(\ref{tc1}) and~(\ref{tc2}), where $G^2 = \frac{d\tau}{dt}$, with $a=0$ and $b=1$ one gets
$\hat a=-k Y$ and $\hat b=\sigma$ proving that $Y(\tau)$ is the Ornstein-Uhlenbeck (OU) process~\cite{Revuz91}:
$$dY = -k Y d\tau + \sigma dW_\tau$$

Note that the OU process is the only stationary process that can be obtained from the Wiener process through a purely deterministic time change. In order to see this, one should consider the equations for $\hat a$ and $\hat b$, imposing $a=0$ and $b=1$ and impose stationarity. The equation for $\hat a$ impose the form of $f(\tau)=e^{\beta\tau}$, while the equation for $\hat b$ determines the form of $t(\tau)$. { Strangely enough, Eq.~\ref{Winer2OUTimeChange} seems not to be very familiar in the statistical physics community.}

\subsection{Example: ABBM/CIR model}

Consider the ABBM model for Barkhausen noise as defined in Eq.~\ref{ABBM}.
The force $F(x)$ is a stochastic term, which behaves as a Brownian process in space.  This means that, if $v$ is non negative, we can write a well defined stochastic equation in terms of the process $v(x)$, where $x$ plays the usual role of time, and write:

$$
dv(x) = k \left( \frac c{v(x)} - 1\right) dx + \sigma dW_x,
$$
where we used the physical relation $v dt = dx$, and $dF(x)=\sigma dW_x$, where $W_x$ is a Wiener process {\em in the space coordinate $x$}.

Now we perform a random time change  $t(x)$ defined by

$$
t(x) = \int _0^{x} \frac{dx'}{v(x')},
$$

which corresponds to Eq.~(\ref{rtc}), where the old time $t$ is now $x$,  the new time $\tau$ is now $t$, and the function of the stochastic process is $g^2(v) = v^{-1}$. Using Eqs.~(\ref{tc1}) and~(\ref{tc2}), the time changed stochastic process reads:

$$dv(t) = \frac{k(\frac{c}v-1)}{v^{-1}} dt + \frac{\sigma}{v^{-1/2}} dW_t,$$

that is exactly Eq.~(\ref{CIRsde}). Such equation defines, in finance, the Cox-Ingersoll-Ross (CIR) model, and it has been proposed as a model for stock prices, first, and than more widely used for describing price volatility~\cite{Jeanblanc2009}.

\section{Exact results for the ABBM-CIR process}\label{CIRSection}

\subsection{Connections with other stochastic processes}

\begin{figure}
\begin{tikzpicture}
  \matrix (m) [matrix of math nodes, row sep=3em, column sep=10em]
    { \text{\bf ABBM}^{(\ref{ABBM})} & \text{\bf CIR}^{(\ref{CIRsde})}  & \text{\bf Rayleigh}^{(\ref{rayleigh})}  \\
      \text{\bf ROU}^{(\ref{ROU})}&\text{\bf GBESQ}^{(\ref{GBESQ})} & \delta\text{\bf-dim. OU}  \\ 
      \text{\bf BES}^{(\ref{BESstandard})}& \text{\bf BESQ}^{(\ref{BESQ},\ref{BESQstd})} &  \delta\text{\bf-dim. BRO}  \\ 
      };
      { [start chain] \chainin (m-2-2);
      \chainin (m-1-2)[join={node[left,labeled]{}}];}
      { [start chain] \chainin (m-3-1);
      \chainin (m-2-1)[join={node[left,labeled]{\text{det. time change}}}];}
        { [start chain] \chainin (m-3-2);
      \chainin (m-2-2)[join={node[left,labeled]{\text{det. time change}}}];}
        { [start chain] \chainin (m-3-3);
      \chainin (m-2-3)[join={node[left,labeled]{\text{det. time change}}}];}
  { [start chain] \chainin (m-1-1);
    \chainin (m-1-2)[join={node[above,labeled]{\text{random time change}}}];
    { [start branch=A] \chainin (m-2-2)
        [join={node[right,labeled] {\delta=\frac{4kc}{\sigma^2}}}];}
    \chainin (m-1-3) [join={node[above,labeled] {\text{Lamperti transform}}}];
    { [start branch=B] \chainin (m-2-3)
        [join={node[right,labeled] {c=\frac{\sigma^2}{4k}\,\,(\text{i.e. }\delta=1)}}];}
}
{ [start chain] \chainin (m-2-3);
    { [start branch=B] \chainin (m-3-3)
        [join={node[right,labeled] {k=0}}];}    
    \chainin (m-2-2) [join={node[above,labeled] {\text{squared mod.}}}];
    { [start branch=A] \chainin (m-3-2)
        [join={node[right,labeled] {k\to 0 \text{ with }\delta\text{ fixed}}}];}
    \chainin (m-2-1)[join={node[above,labeled] {\text{square root}}}];
    { [start branch=A] \chainin (m-3-1)
        [join={node[right,labeled] {k\to 0 \text{ with }\delta\text{ fixed}}}];}
}
  { [start chain] \chainin (m-3-3);
      \chainin (m-3-2) [join={node[above,labeled] {\text{squared mod.}}}];
      \chainin (m-3-1)[join={node[above,labeled] {\text{square root}}}];
    }
\end{tikzpicture}
\caption{Relations between stochastic processes considered in this paper~(see also~\cite{Going-Jaeschke2003}): The ABBM model for Barkhausen noise, the CIR model in finance, and the Bessel processes, as the Bessel process (BESQ) and the Generalised Bessel process (GBES), which recover the squared modulus of a Brownian motion (BRO) and an Ornstein-Uhlenbeck model (OU), for integer dimensionality $\delta$, respectively, as the Bessel process (BES) and its generalised mean reversing version (GBES) do with the modulus of BRO and OU. Labels on the arrows indicate the transformation needed to pass from a model to the other. The numbers near the acronyms refer to the equations in the text. \label{FigStochasticProcesses}}
\end{figure}
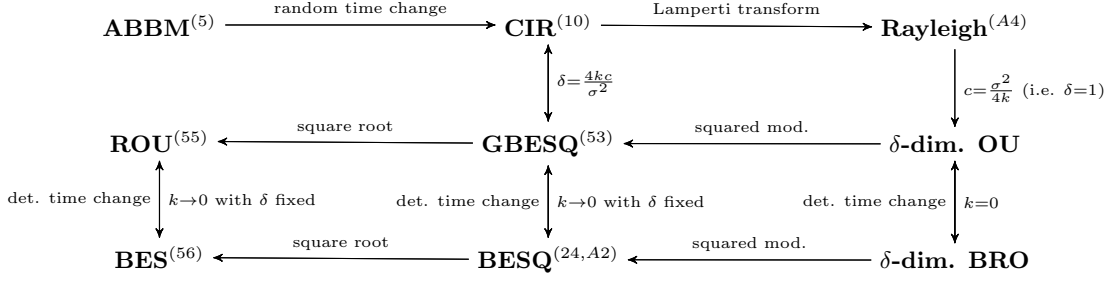

{ Before considering the solution of~\eqref{CIRsde}, we discuss the interesting connections of such equation with some fundamental stochastic processes. In the previous sections we showed that time changes can usefully transform a stochastic equation into an other. For instance, the OU process may be seen as a deterministic time change of a Wiener or Brownian (BRO) process.  We also showed that, using a random time change, the stochastic process defining the ABBM model, which is a zero temperature, driven diffusion in a quenched random environment, is equivalent to a diffusion process with a usual "thermal" noise, but with a multiplicative diffusion coefficient, the CIR process.

An other common way to transform a (multiplicative) stocastic differential equation is via the so called Lamperti transform~\cite{Gardiner1985}, that can get rid of the varying diffusion coefficient via a suitable chosen new stochastic variable and the application of Ito's formula. This can be exploited to further transform the CIR process in the Rayleigh process, which may model the overdamped motion of a Brownian particle in potential consisting in a quadratic well plus a logarithmic correction. In the limit of vanishing logaritmic correction, the Rayleigh process obviously recover the OU.

More interestingly, there appears to be a more fundamental and direct connection of CIR process with OU. If one considers the overdamped diffusion of a Brownian particle in a $\delta$-dimensional space in presence of a square well potential, that is the $\delta$-dimensional version of the OU process, it is quite easy to note that the squared modulus of the distance from the minimum of the well satisfies a stochastic equation (known as the Generalized Squared Bessel process, or GBESQ) which is exactly th CIR stochastic equation, where the dimensionality $\delta$ determines the drive $c$, since 
\begin{equation}
\delta = 4 kc/\sigma^2 \label{deltadefinition}
\end{equation}

Furthermore, exploiting the aforementional deterministic time change, the $\delta$-dimensional OU process, can be recasted in a $\delta$-dimensional BRO process. Via the same time change, the GBESQ recover the Squared Bessel process (BESQ), which represent the squared modulus of a $\delta$-dimensional BRO process. 

Finally, the squared root of GBESQ and BESQ are, respectively, the Generalized Bessel process, also known as the radial OU process (ROU), and the Bessel process (BES).

 We invite the reader interested in the details to Appendix~\ref{connections}, while in Table~\ref{FigStochasticProcesses} we summarize all these connections, which has been previously discovered in the study of stochastic processes~\cite{Going-Jaeschke2003}.}

\subsection{Solution of the ABBM/CIR Model}\label{CIRSolutionSection}

The stochastic process~\eqref{CIRsde} is fully characterized by the following Fokker-Planck (FP) equation (we recall that we are always in the It\^o's integration scheme framework):

\begin{equation}
\partial_t P(v,t;v_0) = -\partial_v \left(k(c-v) P(v,t;v+0)\right) + \frac 12 \partial^2_{v} \left( \sigma^2 v P(v,t;v_0) \right) = -\partial_x J(v,t;v_0) \label{FPABBM}
\end{equation}

where $P(v,t;v_0)$ is the probability that the process takes value $v$ at time $t_0+t$, given that it took value $v_0$ at time $t_0$, and, correspondingly
\begin{equation}
J(v,t;v_0) \equiv  \left(k(c-v) P(v,t;v_0)\right) - \frac 12 \partial_{x} \left( \sigma^2 v P(v,t;v_0) \right) \label{current}
\end{equation}
 is the probability current.

Firstly we consider the stationary solution of the equation, that is a normalizable distribution $P_s(v)$ with null probability current:
\[
 \left(k(c-v) P_s(v)\right) - \frac 12 \partial_{v} \left( \sigma^2 v P_s(v) \right)=0
\]
It is easy to verify that, naming   $\delta \equiv  \frac{4kc}{\sigma^2}$,

\begin{equation}
P_s(v) = \frac{\left(\frac{2c}{\delta}\right)^{-\frac{\delta}{2}}}{\Gamma\left(\frac{\delta}{2}\right)} v^{ \frac{\delta}{2} -1}e^{-\frac {2\delta}{c} v},\label{CIRstationary}
\end{equation}
which is a Gamma distribution with shape parameter $\frac{2kc}{\sigma^2}=\frac{\delta}2$ and rate $\frac{2k}{\sigma^2}=\frac{2\delta}c$.

Before proceeding with the exact analysis of the non stationary solutions of the FP equation, let us informally discuss what we expect.

{ The non stationary solution of the FP equation is the propagator of the continuous Markov process $P(v,t;v_0,t_0)$, and represents the probability that a single stochastic trajectory started from $v_0$ at time $t_0$ will reach $v$ at time $t>t_0$. The solution of the FP equation requires some boundary conditions. Usually, two possible different boundary conditions are chosen~\cite{Gardiner1985}: 
\begin{enumerate}
  \item Reflecting
  \item Absorbing.
\end{enumerate}
Reflecting boundary conditions correspond to a solution with zero current at the border. This assures that the total probability, that is the norm, or the "mass", keeps constant during the dynamics:
\[
\int P(v,t;v_0,t_0) dv = 1.
\]
This solution represents the trajectories that freely evolve in the domain of the solution, and rebound, i.e. are reflected, at the border.

On the other hand, absorbing boundary conditions are usually obtained imposing a vanishing propagator at the border
\[
\left.P(v,t;v_0,t_0)\right|_{\text{for $v$ at the border}} =0.
\]
Usually this also guarantees that the propagator vanishes for initial condition at the border 
\[
\left. P(v,t;v_0,t_0)\right|_{\text{for $v_0$ at the border}}= 0.
\]
 This solution describes trajectories that are absorbed or "killed" when they touch the border, since they can not proceed further: the probability to start from the border and going elsewhere is always zero.

This common recipe, however, does not apply in the case in study here. The border of interest here, that is $v=0$, is special, since the diffusion coefficient is zero there. This means that, the current in Eq.~\eqref{current} at the border $v=0$ should read (provided that  $P(v,t;v_0,t_0)$ is not singular in $v=0$):
\[
k\, c \,P(0,t;v_0,t_0),
\]
and this shows that reflecting and absorbing boundary conditions should coincide, which does not easily agree with our intuition of the meaning of the corresponding solutions.

The fact is that the usual recipe for obtaining the desired solution, only applies for "regular" boundaries, that is where the drift and diffusion coefficients of the FP are not zero (or singular). In the other case, that is for the so called "natural" borders, one can not simply impose arbitrary boundary conditions. As explained below, one should find a solution and then compute its behaviour at the border in order to understand its meaning in terms of trajectories. 

As we will see in the following, our FP equations has only two acceptable solutions as propagators for a Markov process. One of this, which we will call $P_+$, diverges at $v=0$ for certain values of the parameters. Nevertheless the limiting value of the current turns to be zero at the border (as a bona fide "reflecting" solution). The other solution, that we will note as $P_-$,  keeps a constant, positive value at $v=0$, but is turns to be zero for $v_0=0$, as an genuine "absorbing" solution should do.

The solution $P_+$ and $P_-$ are the main tools for the derivation of all the main results of the present work. In the following we will describe their derivation and the main features and differences.}

\subsection{Propagator via time change}

The (reflecing) propagator of the CIR process, in the context of ABBM model, has been computed by Bertotti, using standard eigen-function technique~\cite{Bertotti1998}. However, the same result can be obtained exploiting the time change transformation described above.

As explained before, the Ornestein-Uhlenbeck process 
$$
dy = -k \,y\, dt +\sigma \,dW_t
$$
can be recast in terms of a time changed Brownian process:
\begin{equation}
y(t)=e^{-\frac k2 t} z\left( \frac{\sigma^2 (e^{kt}-1)}{4k}\right) \label{tcW2OU},
\end{equation}
where 
\begin{equation}
dz = \sigma \,dW_t \label{sdeBr}
\end{equation}
whose propagator is a Gaussian distribution
$$
P_{Bro}(z,t;z_0) = \sqrt{\frac{1}{2\pi \sigma^2 t}} \exp\left(-\frac{(z-z_0)^2}{2 \sigma^2 t}\right).
$$
Accordingly to Eq.~(\ref{tcW2OU}), the propagator for the Ornstein-Uhlenbeck process is:
$$
P_{OU}(y,t;y_0) = P_{Bro}\left(e^{\frac k2 t} \,y, \frac{\sigma^2 (e^{kt}-1)}{4k};y_0\right) e^{\frac k2 t} 
$$
that is:
$$
P_{OU}(y,t;y_0) = \sqrt{\frac{k}{2 \pi (1-e^{-kt})\sigma^2}} \exp\left[-\frac{k(y-y_0 e^{-\frac k2 t})^2}{2(1-e^{-kt})\sigma^2 }\right]
$$

Let us now consider a generic Brownian motion in dimension $\delta$. More specifically we consider a stochastic process in $\delta$ dimension, where each coordinate evolve independently according to the same sde Eq.~(\ref{sdeBr}).
Now, we are interested in the square of the modulus of such a $\delta$ dimensional Brownian process:

$$
w = \sum_{i=1}^\delta z_i^2
$$
The process satisfies the following sde:
\begin{equation}
dw = \frac{\delta \sigma^2}4  dt + \sigma \sqrt{w} \,dW_t\label{BESQ}
\end{equation}
which is the square of a Bessel process.
Its propagator is given by:
$$
P_{BESQ}(w,t;w_0) = \frac{2}{\sigma^2 t} \left(\sqrt{\frac{w}{w_0}}\right)^{\frac\delta2-1}\exp\left(-2\frac{w+w_0}{\sigma^2 t}\right) I_{\frac\delta2-1}\left(4\frac{\sqrt{w\,w_0}}{\sigma^2 t}\right)
$$
where $I_n$ is the modified Bessel function of the first kind and of order $n$:
\[
I_n(x) = \sum_{r=0}^{\infty} \frac{(x/2)^{2r + n}}{r!\Gamma(r+1+n)}.
\]
The propagator $P_{BESQ}$ is known as a  non-central $\chi^2$-distribution. If we now perform, on each coordinate of the $\delta$ dimensional Brownian process, the same time change relation Eq.~(\ref{tcW2OU}), we obtain a $\delta$ dimensional OU processes. 

Accordingly, the squared modulus of the two $\delta$ dimensional process, will be related by the time change transformation:
$$
v=e^{-kt} w\left( \frac{\sigma^2 (e^{kt}-1)}{4k}\right).
$$

Since the corresponding sde for $v$ is the CIR equation, one has that~\cite{Going-Jaeschke2003,Jeanblanc2009}

$$
P_{CIR}(v,t;v_0) = P_{BESQ}\left(e^{kt}v, \frac{\sigma^2 (e^{kt}-1)}{4k};v_0\right) e^{kt},
$$

and setting $c = \frac{\delta\sigma^2}{4k}$, we obtain the CIR propagator:

\begin{equation}
P_{CIR}(t,v;v_0) = \lambda \exp\left[-\lambda\cdot(v+v_0 e^{-k t})\right] \cdot \left(\frac{v}{v_0 e^{-k t}}\right)^{\mu/2} \cdot I_{\mu}\left(2\lambda\sqrt{v\,v_0 e^{-k t}}\right),\label{CIRprop}
\end{equation}
where
\[
\lambda(t) \equiv \frac {2 k}{\sigma^2 \left(1-e^{-k t}\right)}
\]
(whenever possible, we will omit the time dependence in $\lambda$, to make formulas lighter), and
\[
\mu \equiv \frac{2kc}{\sigma^2} -1 = \frac {\delta}{2} -1.
\]

The solution Eq.~(\ref{CIRprop}) is the propagator of the CIR/ABBM process, since it satisfy the properties
$P_{CIR}\ge0$, $\lim_{t\to0} P_{CIR}(v,t;v_0)=\delta(v-v_0)$ and
$$
\int_0^\infty P_{CIR}(v,t;v_0) dv = 1
$$
However, how we will recall below, this is not the only positive solution of the FP equation Eq.(\ref{FPABBM}), if we relax this last property.

\subsection{First return of Brownian motion and dynamical regimes of CIR process}

Knowing the relation between the CIR model and the square modulus of a Brownian motion, we can immediately understand a main feature of its dynamics. Indeed, we know that the statistics of zero passages of Brownian motion change with dimensionality, and in particular that for $\delta < 2$ the process is recurrent (i.e. it returns to the origin almost surely), while for $\delta\ge 2$ it is not. This reflects on the properties of the CIR model for $c < \frac{\sigma^2}{2k}$,  and $c>\frac{\sigma^2}{2k}$ respectively. In fact, the index $\mu$ of the Bessel function in the CIR propagator Eq.~(\ref{CIRprop}) change signs, and this changes its behaviour for small argument $x\approx 0$, which is 
\begin{equation}
I_\mu(x) \approx  \frac{(x/2)^{\mu}}{\Gamma(1+\mu)} + O(x^{2+\mu})\label{Bsmall}.
\end{equation}
The case $\delta\le2$ corresponds to $\mu\le0$, while $\delta>2$ to $\mu>0$. 

{ As we mentioned before, the drift coefficient of CIR equation vanishes at $v=0$, making this border a "natural border" of the stochastic process. This means that one cannot use the standard recipes to impose reflecting or absorbing boundary conditions at will. The theory of stochastic processes has studied all possible cases of natural borders for one dimensional processes, classifying their respective behaviours, even if there are different nomenclatures adopted in the literature. In the specific case of CIR, the classification of the border $v=0$ depends on the value of $\mu$ (or $\delta)$). In particular, the border  changes from {\em entrance-exit} or {\em non-singular}~\cite{Jeanblanc2009} for $-1<\mu<0$ (or {\em Regular} in the Feller's classification, or {\em Regular-Attracting-Attainable} for the Russian school~\cite{Karlin1981}),  to {\em entrance-not exit}~\cite{Jeanblanc2009} for  $\mu\ge 0$ ({\em Entrance} for Feller, {\em Natural-Non attracting-Unattainable} for the Russian school~\cite{Karlin1981}).

In practice, this means that for $\mu<0$  ($\delta<2$) the process do touch (infinitely many times) the value $v=0$, keeping neverthelss its non negative value, while for $\mu>0$ ($\delta>2$) the process has zero probability to attain the $v=0$ value (but for $\mu = 0$ it can return arbitrarily close with probability one)~\cite{Redner2001}.}

\subsection{Feller's solutions}

To our knowledge, the first study of Eq.~(\ref{FPABBM}) is a paper by Feller~\cite{Feller1951}, who finds all the solutions of the equation for the general values of the parameters $k, c, \sigma$. Here we discuss the acceptable solutions for $c,k>0$ (i.e. $\mu>-1$, or $\delta>0$) valid as propagators of the stochastic process, i.e. functions $P(v,t;v_0)\ge 0$ such that the norm remains finite:
\begin{equation}
M(t)  \equiv \int_0^\infty P(v,t;v_0)dv\le 1,\label{mass}
\end{equation}
and $\lim_{t\to 0} P(v,t;v_0) = {\delta}(v-v_0)$ (here $\delta$ is the Dirac distribution). 

As previously mentioned, the relevant parameter of the equation is $\mu$, since it embodies the recurrence properties of the diffusion process and discriminates between two different dynamical behaviours for $\mu<0$ and $\mu>0$, which corresponds to slow $0 < c < \sigma^2/2k$ or fast driving $c>\sigma^2/2k$, respectively, or to $\delta<2$ and $\delta>2$, in terms of OU-dimensionality.

\subsubsection{"Crackling" or "stick-slip" behaviour: $-1<\mu<0$ (i.e. $0 < 2 k c < \sigma^2$ or $\delta<2$)} 
In this case there are only two acceptable solutions, which differs for the time evolution of the norm:
\[
M(t) =\int_0^\infty P(v,t;v_0) dv
\]
\begin{enumerate}
\item A "reflecting" solution, with zero current in $v\to 0^+$, whose norm is constant (we choose the natural value $1$):
$M(t) = 1$.  Feller does not provide the explicit expression for the "reflecting" solution, but this is nothing else than the CIR propagator given above:
\begin{equation}
P_{+}(v,t;v_0) \equiv P_{CIR}(v,t;v_0)=\lambda \exp\left[-\lambda\cdot(v+v_0 e^{-k t})\right] \cdot \left(\frac{v}{v_0 e^{-k t}}\right)^{\mu/2} \cdot I_{\mu}\left(2\lambda\sqrt{v\,v_0 e^{-k t}}\right).\label{reflecting}
\end{equation}
Interestingly, in~\cite{Dornic2005} the authors note that $P_+$ can be written also as:
\begin{equation}
P_+(v,t;v_0) = \sum_{n=0}^\infty \frac{\left(\lambda v_0 e^{-k t}\right)^n \exp\left(-\lambda v_0 e^{-k t}\right)}{n!} \frac{\lambda e^{-\lambda v}\left(\lambda v\right)^{n+\mu}}{\Gamma(n+\mu+1)} \label{mixture}
\end{equation}
This expression shows that $P_+$ is a mixture of two elementary distribution functions: the Gamma distribution
$\Gamma_{\alpha,\beta}(y) = \beta^\alpha y^{\alpha-1} e^{-\beta y}/\Gamma(\alpha)$, where $\alpha$ and $\beta$ are, respectively, the shape and the rate parameter, and  the Poisson distribution $P_\nu(n)=\nu^n \exp(-\nu)/n!$.  In other words, $P_+$ is a Gamma distribution with a Poisson distributed shape parameter. This observation can be useful for numerical computations. { Note also, that the expression Eq.~\ref{mixture} is very handy to check that $\lim_{t\to\infty} P_+(v,t;v_0) = P_s(v)$, since in this limit, only the $n=0$ term of the sum survives.}

\item An "absorbing" solution, with finite, non zero current in $v=0$ and norm which decreases to zero $\lim_{t\to \infty} M(t) = 0$. It reads
\begin{equation}
P_{-}(v,t;v_0) = \lambda \exp\left[-\lambda\cdot(v+v_0 e^{-k t})\right] \cdot \left(\frac{v}{v_0 e^{-k t}}\right)^{\mu/2} \cdot I_{-\mu}\left(2\lambda\sqrt{v\, v_0 e^{-k t}}\right)\label{absorbing}.
\end{equation}

Note that this solution reaches a finite, not null value, for $v\to 0^+$, while it vanishes for $v_0\to 0$. The expression for $P_{-}$ is identical to $P_{CIR}=P_+$, but for the index of the Bessel function, which has the opposite sign.

\end{enumerate}
 
We now discuss in detail the differences between the "reflecting" $P_+$ and the "absorbing" $P_-$ solutions.
\begin{enumerate}
\item Firstly, the two distribution differs for their norms: the reflecting solution as a constant norm (this can be easily seen from the mixture expression~\eqref{mixture}): correspondingly, the current probability is null at $v=0$. On the other hand, the "absorbing" solution has a decreasing norm~\cite{Feller1951}:
\begin{equation}
\int_0^\infty P_-(v,t;v_0) dv = \Gamma\left(-\mu;\lambda(t) v_0 e^{-k t}\right) \label{massabs}
\end{equation}
where $\Gamma(n;z)=\frac 1{\Gamma(n)}\int_0^z e^{-x} x^{n-1}dx$. Note that for $t\to 0$, $\exp(-k t) \lambda(t) \to \infty$ and the norm goes to $1$ (since $\Gamma(n;\infty)=1$). The norm is always decreasing for large $t$, and it goes to zero for $k>0$. 
\item
Secondly, the solutions differ for the behaviour for $v$ or $v_0$ small, that is for $v\,v_0$ small. In order to see this, let substitute to the Bessel function its dominant behaviour for its argument  (see Eq.~(\ref{Bsmall})), obtaining:
\begin{equation}
P_{\pm}(v,t;v_0) = \frac{\lambda^{1\pm\mu}}{\Gamma(1\pm\mu) }\exp\left[-\lambda\cdot (v+v_0 e^{-k t})\right] 
\cdot \left(\frac{v}{v_0 e^{-k t}}\right)^{\mu/2} \left[ \lambda^{\pm\mu}e^{-(\pm \mu kt)/2}\left (v\,v_0\right)^{\pm\mu/2} + O\left([v\,v_0]^{1\pm\mu/2}\right)\right]\label{usmall}.
\end{equation}

So, the dominant behaviour at small $v = \epsilon \approx 0$ is:
\begin{eqnarray}
P_+ (\epsilon,t;v_0)&= & \frac{\lambda^{1+2\mu}}{\Gamma(1+\mu) }\exp\left[-\lambda\cdot(v_0 e^{-k t})\right] \cdot \epsilon^{\mu} +  O\left(\epsilon^{1+\mu}\right)\label{Rsmallx}\\
P_-(\epsilon,t;v_0) &=& \frac{\lambda^{1-2 \mu}}{\Gamma(1-\mu) }\exp\left[-\lambda\cdot(v_0 e^{-k t})\right] \cdot  
v_0^{-\mu}e^{\mu k t} + O\left(\epsilon\right).\label{Asmallx}
\end{eqnarray}
Recalling that we are studying the case $-1<\mu<0$ we see that the solution have a different behaviour for $v\approx 0$, since $P_+$ diverges, while $P_-$ remains finite.
On the other hand, for small $v_0=\epsilon$:
\begin{eqnarray}
P_+ (v,t;\epsilon)&= & \frac{\lambda^{1+\mu}}{\Gamma(1+\mu) }\exp\left[-\lambda v \right] \cdot v^{\mu} + O\left(\epsilon\right) \label{Rsmallxi}\\
P_-(v,t;\epsilon) &=& \frac{\lambda^{1-\mu}}{\Gamma(1-\mu) }\exp\left[-\lambda v \right] \cdot  
 \epsilon^{-\mu}e^{\mu k t} + O\left(\epsilon^{-\mu}\right) \label{Asmallxi},
\end{eqnarray}
which means that even the behaviour for $v_0 \approx 0$ is different: now $P_-$ goes to zero, while $P_+$ remains finite. 
\end{enumerate}
This observation justify the naming of $P_+$ and $P_-$ as reflecting and absorbing, respectively: $P_+$ describes trajectories which goes arbitrarily close to zero and leaves zero with probability one; $P_-$ describes trajectories that, once they touch $v=0$ are absorbed (the probability to start a trajectory from zero is zero). Let observe here that the "reflecting/absorbing" terminology is not obvious in this problem, due to the singular (vanishing)  nature of the diffusion coefficient in the Fokker-Planck equation. In these cases, we can not impose arbitrary boundary conditions as in the more regular cases.

\subsubsection{"Steady-sliding" behaviour: $\mu>0$ (i.e. $2 k c > \sigma^2$ or $\delta>2$)}

For the fast driving case $c>\sigma^2/2k$, Feller shows that there's a single acceptable solution that has a constant norm (zero probability current in $v=0$): this means that this solution is "reflecting", despite the fact that it vanishes for $v\to 0^+$. The solution reads:
\begin{equation}
P_1(v,t;v_0) \equiv \lambda \exp\left[-\lambda (v+v_0 e^{\-k t})\right] \left(\frac{v}{v_0 e^{-k t}}\right)^{\mu/2} I_{\mu}(2\lambda\sqrt{v\,v_0 e^{-k t}})
\end{equation}

The solution of the equation for $\mu>0$ is well known in finance, since it has been proposed as a model for the interest instantaneous rate.

Note that this solution is identical to the "reflecting" solution $P_1 = P_+$, but now it is not diverging for $v=\epsilon\approx 0$, since $\mu>0$:
\begin{equation}
P_1(\epsilon,t;v_0) \approx \frac{\lambda^{1+\mu}}{\Gamma(1+\mu)}\exp\left(-\lambda v_0 e^{-k t}\right) \epsilon^\mu.
\end{equation}

\subsubsection{Threshold value: $\mu=0$ (i.e. $2 k c = \sigma^2$ or $\delta=2$)}
For $\mu=0$ (that is $2kc=\sigma^2$ or $\delta=2$), all the previous solution coincide $P_+=P_-=P_1$, and read:
\begin{equation}
P(v,t;v_0) \equiv \lambda \exp\left[-\lambda (v+v_0 e^{\-k t})\right] I_{0}(2\lambda\sqrt{v\,v_0 e^{-k t}}).
\end{equation}
This propagator has constant norm, so the "absorbing" solution is lost.
Moreover, since the small argument $x\approx 0$ behaviour of the Bessel function with index zero is:
\[
I_0(x) = 1 + O(x^2)
\]
the solution goes to a finite constant, both for $v\approx 0$, as well as for $v_0\approx 0$.

{In the Appendix~\ref{generating} we provide other known results, as the  expression (whatever the value of $\mu$) of several relevant averages:
\[
\langle v^m x^n\rangle_{v_0},
\]
where $x = \int_0^t v(s)ds$ is the displacement at time $t$ and the average is over the (reflecintg) solution with initial condition $v(0)=v_0$.}

\section{Avalanche (excursion) and multi-avalanche (bridge) shapes in the ABBM/CIR/Bessel process} \label{ExcursionBridgeSection}

\subsection{Avalanches in ABBM/CIR process}

The ABBM model has been proposed to describe the intermittent properties of Barkhausen noise. This can be characterised by the statistics of avalanches, that is the portion of the signal comprises between two consecutive passage through a small threshold value, approximately close to zero. In the theory of stochastic process, this quantity is known as the excursion of the process, which, in this case, is a trajectory starting from zero and returning to zero {\em for the first time} after a time $T$. 

\subsubsection{Distribution of durations}

{The probability to observe an avalanche can be estimated by the distribution of durations. In the original treatement of the ABBM model, the distribution of durations has been informally derived from the stationary distribution Eq.~\eqref{CIRstationary} plus a scaling hypothesis for its power law regime, obtaining an algebraic decay of $P(T)$ with an exponent varying linearly with the drive, from $-2$ (for vanishing drive) to $-1$ just before the steady sliding phase~\cite{Bertotti1998}.
Such result, fairly confirmed by experiments, is quite striking, since in general the first return distributions (such as the $P(T)$) are not expected to be simply related to the stationary distribution, which is the asymptotic limit of a one-time quantity (at odds with first return distribution, which involve a contraints on an continuum of times).
Nevertheless, the algebraic $-2$ decay has been recovered exactly in the limit of vanishing drive $c\to 0$ (or $\mu\to -1$ in our notation), in~\cite{LeDoussal2012}, and in the case of a small "kick", that is for the avalanche following a non zero driving field applied for an infinitesimal time interval.

Here we provide the exact distribution of durations for a generic drive $\mu<0$. }This quantity can be obtained from the expression of the norm for the "absorbing" solution, Eq.~(\ref{absorbing}) (we obviously are restricting to the $\mu<0$ case, where the absorbing solution exists). Considering the probability that a trajectory started from a small value $\epsilon$ has survived until time $T_\epsilon$, the cumulative distribution of $T_\epsilon$ is:
\begin{eqnarray}
\text{Prob}(T_\epsilon>T) &=& \int_0^\infty P_-(v,T;\epsilon)dv.
\end{eqnarray}

Using the expression for small $\epsilon$, Eq.(\ref{Asmallxi}), it reads:
\[
\text{Prob}(T_\epsilon>T) \approx \frac{\epsilon^{-\mu}}{\Gamma(1-\mu)} \left[\lambda(T)e^{-kT}\right]^{-\mu}.
\]

Note that the probability is zero if $\epsilon=0$, since it is computed using the absorbing propagator, which is null if $v_0=\epsilon\to0$. Strictly speaking, as for the Wiener process, the probability of first re{}turn to zero is a singular distribution and $P(T_0<T)=1$ for every $T>0$, since the process returns infinitely many times in every interval $(0,T)$. 
Nevertheless, in order to consider the excursions of finite durations, we can consider a small but positive starting value $\epsilon>0$. Then, since $T \ll 1/k$, 
$\lambda(T)e^{-kT} \approx \frac{2}{\sigma^2 T}$,
and we have
\[
\text{Prob}(T_\epsilon>T) \approx  \left[\frac{2\epsilon}{\sigma^2}\right]^{-\mu} \frac{T^{\mu}}{\Gamma(1-\mu)},
\]
so the distribution of duration is, for $T\ll 1/k$:
\[
P(T) = -\frac{d}{dT} \text{Prob}(T_\epsilon>T) \approx \left[\frac{2\epsilon}{\sigma^2}\right]^{-\mu} \frac{T^{\mu-1}}{\Gamma(-\mu)},
\]
where we used $\Gamma(1-\mu)=-\mu \Gamma(-\mu)$.

In the original parameters of ABBM model, this regime holds for $T\ll 1/k$:
\[
P(T) \propto T^{-2+\frac{2kc}{\sigma^2}}
\]
while it decrease exponentially for $T\gg 1/k$. This recovers the expected decay, with a drive depending exponent, as predicted before~\cite{Bertotti1998}. 
Moreover, note that, since $\frac{2kc}{\sigma^2} = \frac{\delta}2$, one obtains for $\delta=1$ the well known $T^{-3/2}$ decay for the return to the origin of one-dimensional Brownian process, whereas the formula can not be applied for $\delta>2$, since this corresponds to $\mu>0$, where the absorbing solution expires.

\subsubsection{Avalanche shape distribution}

Now we consider the shape of the avalanche, that is the probability distribution that, during an avalanche, the process takes a value $v$ at time $0<t<T$, where $T$ is the duration of the avalanche.
Such distribution can be expressed, given the knowledge of the "absorbing" propagator $P_-$, as:
\begin{equation}
P_E(v,t;T) = \lim_{\epsilon\to 0} \frac{P_-(v,t;\epsilon) P_-(\epsilon,T-t;v) }{\int_0^\infty P_-(u,t;\epsilon) P_-(\epsilon,T-t;u) du}\label{pexc},
\end{equation}
which is the probability that, starting at $t=0$ from a value $\epsilon$ arbitrarily close to $0$, reaches the value $v$ at time $t$, and then comes back to $\epsilon$ in a time $T-t$. The denominator realizes the correct normalization of the probability and the use of the absorbing propagator $P_-$ make sure that we are counting only the trajectories that do not touch zero during their promenade. This distribution allows to compute, for instance, the average shape of the avalanche:
\[
\langle v(t)\rangle_E = \lim_{\epsilon\to 0} \int_0^\infty v \cdot P_E(v,t;T) dv\label{avex},
\]
where $0<t<T$ is the internal time inside the avalanche.
The computation of $P_E(v,t;T)$ is based on the determination of the kernel:
\[
K_\epsilon(v) =  P_-(v,t;\epsilon) P_-(\epsilon,T-t;v).
\]
Using the expressions for small $\epsilon$ Eqs.~(\ref{Asmallx}) and~(\ref{Asmallxi}), one gets:
\[
K_\epsilon(v) \approx \left\{
		\frac{\lambda(t)^{1-\mu}}{\Gamma(1-\mu) } \exp\left[-\lambda(t) v \right] \cdot  
 		\epsilon^{k}e^{\mu k t/2}
		\right\}
		 \left\{ 
		 	\frac{\lambda(T-t)^{1-\mu}}{\Gamma(1-\mu) }\exp\left[-\lambda(T-t)\,(v e^{-k (T-t)})\right] \cdot  
			v^{-\mu}e^{\mu k t/2}
		\right\} + O\left(\epsilon\right).
\]
However, since every factor that does not depend on $v$ also appears in the denominator of Eq.~(\ref{pexc}), $P_E$ can be computed using an equivalent kernel, which happens to be also independent on $\epsilon$:
\[
{\cal K}(v) \equiv  \exp\left[-\omega(t,T) v\right] v^{-\mu}
\]
where
\begin{equation}
\omega(t,T) \equiv \left[\lambda(t)+\lambda(T-t) e^{-k(T-t)}\right]= \frac{2k(1-e^{-k T})}{\sigma^2 (e^{-k t}-1)(e^{-k (T-t)}-1)}.
\label{omega}
\end{equation}

In other words, the shape distribution Eq.(\ref{pexc}) is nothing but a simple Gamma distribution (see Figs.~\ref{fig2} and~\ref{fig3}, left panels):
\begin{equation}
P_E(v,t;T) = \frac {{\cal K}(v)}{\int_0^\infty {\cal K}(u) du} = \frac{\omega(t,T)^{1-\mu}}{\Gamma(1-\mu)}\ e^{-\omega(t,T) v} v^{-\mu},\label{ExcDist}
\end{equation}
with shape $1-\mu$ and rate $\omega(t,T)$, whose moments are:
\[
\langle v(t)^n \rangle_E = \frac{(n-\mu)(n-1-\mu)...(1-\mu)}{\omega(t,T)^n}=\frac{\Gamma(n+1-\mu)}{\Gamma(1-\mu)} \omega(t,T)^{-n}.
\]
For instance the average shape of the avalanche is:
\begin{equation}
\langle v(t) \rangle_E = \frac{1-\mu}{\omega(t,T)} = \frac{(1-\mu) \sigma^2}{2 k } \frac{(e^{-k t}-1)(e^{-k (T-t)}-1)}{1 - e^{-k T}}
\label{CIRavex}
\end{equation}
(see panel c in Fig.~\ref{fig1}).

A number of observations are in order:
\begin{enumerate}
\item The shape of the avalanche depends on $\mu$, that is from the drive $c$, just for its global amplitude: the time dependence of the {\em normalised average shape} (for instance the shape divided by its area or by its maximum value) does not depend on $\mu$, and hence on the drive $c$.
\item
The shape can be expressed using hyperbolic functions (recovering previous results~\cite{LeDoussal2012} and~\cite{Papanikolaou2011}), using:
\[
(e^{-k t}-1)(e^{-k(T-t)}-1)  = 4 e^{\frac{-k T}2}  \sinh\left[\frac{k t}{2}\right]\sinh\left[\frac{k (T-t)}2\right]
\]
In~\cite{Papanikolaou2011} and~\cite{LeDoussal2012} this result is obtained in the limit of vanishing drive $c\to 0$ or $\mu\to -1$.
\item
If we consider the rescaled internal time $\tau=t/T$, then for $k T\ll 1$ and fixed $\tau$, the shape takes a parabolic form~\cite{Papanikolaou2011}:
\[
\sinh\left[\frac{k T \tau}2\right] \sinh\left[\frac{k T(1-\tau)}2\right] \approx (k T)^2\,\left[\frac{\tau(1-\tau)}4\right]
\]
A generalization of this parabolic average shape, encompassing the semicircular shape of the Brownian excursion~\cite{Baldassarri2003} as well as the shape of Levy processes~\cite{Colaiori2004}, but also including asymmetric shapes, has been proposed~\cite{Laurson2013} in terms of a universal scaling function, observed in several different phenomena.

\item For $T\to \infty$ the rate function of the excursion Gamma distribution becomes:
\[
\lim_{t\to\infty} \omega(t,T) = \frac{2k}{\sigma^2\left(1-e^{-kt}\right)} = \lambda(t)
\]
and the distribution represents the probability of the infinite meander (see later)

\end{enumerate}

\subsection{Multi-Avalanches in ABBM/CIR process}

\subsubsection{Probability of a multi-avalanche (bridge)}

The probability to observe a multi-avalanche of duration $T_B$ between $T$ and $T+\Delta T$ can be computed using both propagators $P_+$ and $P_-$. The simplest reasoning is the following: first one considers the trajectories that go from $v=0$ at $t=0$ up to a value $v=u$ at $t=T>0$, without any restriction. Then, one consider the probability of the trajectory that start at $u$ and crosses $v=0$ in an interval $dT$. That is:
\[
\text{Prob}\left(T<T_B<T+\Delta T\right)=\int_0^\infty P_+(u,T;0) \left[1-\int_0^\infty P_-(w,\Delta T;u) dw\right] du
\]
Using Eq.~(\ref{massabs}) and Eq.~(\ref{Rsmallxi}) one gets:
\[
 \text{Prob}\left(T<T_B<T+\Delta T\right)=\int_0^\infty  \frac{y^{\mu}e^{-y}}{\Gamma(1+\mu) }\cdot\left[\int^{\infty}_{  y_0}\frac{x^{-\mu}e^{-x}}{\Gamma(-\mu)}dx\right]\,dy,
\]
where $y_0\equiv\frac{\lambda(\Delta T)}{\lambda(T)}y$. 

Since $\lambda(\Delta T)\approx \frac{2}{\sigma^2 \Delta T}$ for $\Delta T\ll 1/k$, using the asymptotic expansion of the incomplete Gamma function~\cite{AbramowitzStegun}:
\[
\int_z^{\infty} e^{-t}t^{a} = z^{a}e^{-z}\left[1+\frac{a}z+\frac{a(a-1)}{z^2}+...\right]\text{ for $z\to\infty$}
\]
our probability becomes at the leading order in $\Delta T$:
\[
 \text{Prob}\left(T<T_B<T+\Delta T\right)\approx \int_0^\infty dy\, \frac{\left(y^{\mu}e^{-y}\right)\left(y_0^{-\mu}e^{-y_0}\right)}{\Gamma(1+\mu)\Gamma(-\mu) } = \frac{k^\mu\,\Delta T^{-\mu}}{\Gamma(1+\mu)\Gamma(-\mu)}\left(1-e^{-k T}\right)^{-\mu}.
\]
{Note that the duration for the multi-avalanche has a completely different behaviour with respect to avalanche durations. For $0<T\ll 1/k$ the probability goes as
\[
\text{Prob}\left(T<T_B<T+\Delta T\right)\approx \frac{\Delta T^{-\mu}}{\Gamma(1+\mu)\Gamma(-\mu)}T^{-\mu}\text{ (for $\Delta T\ll 1/k$ and $0<T\ll 1/k$)} ,
\]
that is it increases with $T$ (remember that $\mu<0$) and goes to the asymptotic value, for $T\gg1/k$
\begin{equation}
  \text{Prob}\left(T<T_B<T+\Delta T\right)\approx \frac{k^{\mu}\Delta T^{-\mu}}{\Gamma(1+\mu)\Gamma(-\mu)} \text{ (for $\Delta T\ll 1/k$ and $T\gg 1/k$)}.\label{AsymptBridgeDuration}
\end{equation}}

\subsubsection{Multi-Avalanche shape}

\begin{figure}[ht]
  \includegraphics[width=12.0cm]{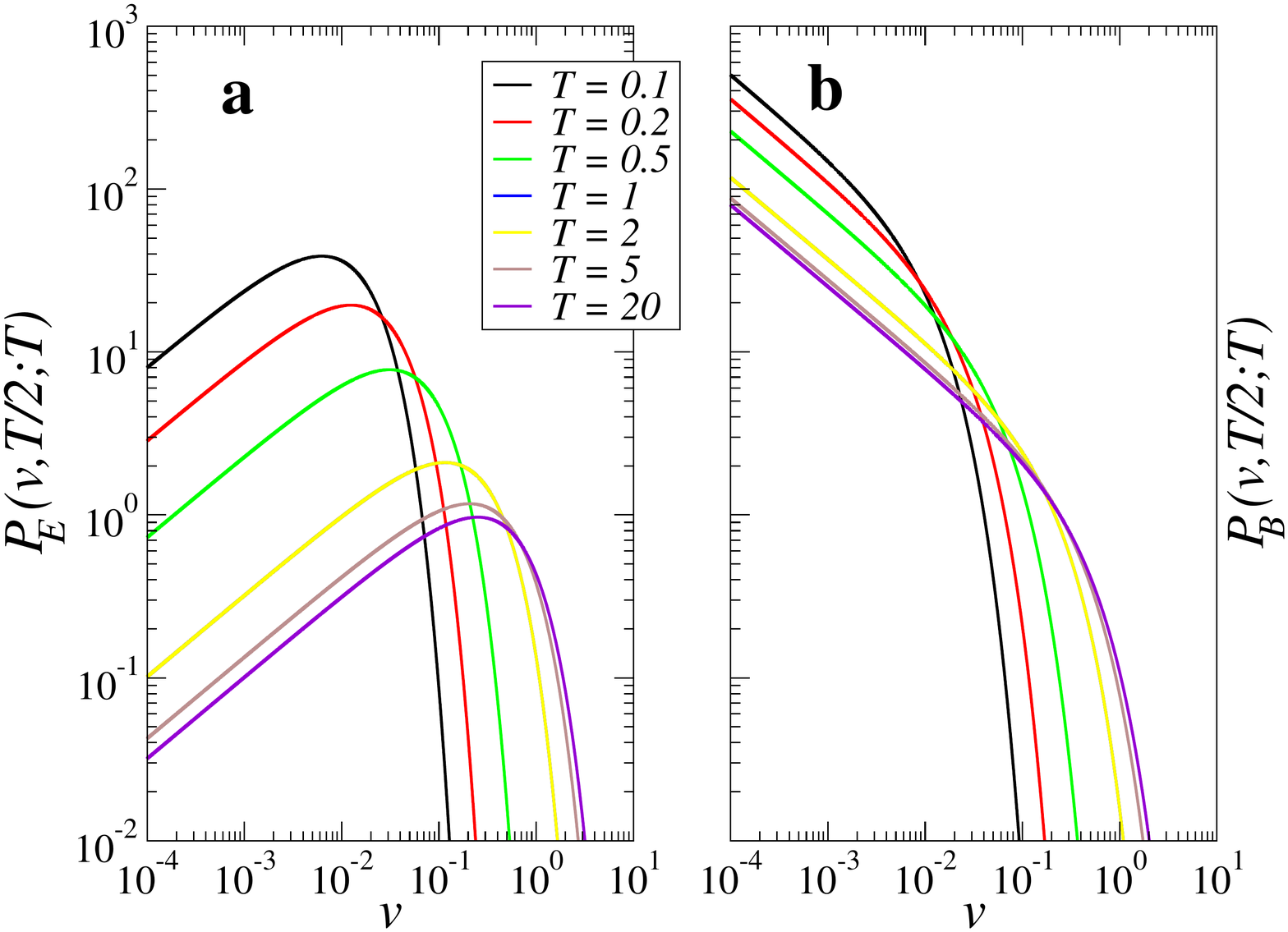}
   \caption{
  Distribution of excursion $P_E(v,t;T)$ (a) and bridge $P_B(v,t;T)$ (b) shape, reported in Eqs.~(\ref{ExcDist}), and~(\ref{BriDist}). The distributions are computed for $\mu=-0.5$, $k=1$, and $\sigma=1$, for $t=T/2$ and different durations $T$. Note that the distributions (both are gamma distributions) differs for the algebraic initial part. For the excursion the probability increase, from $v=0$, and attains a most probable value, while for the bridge, the distribution is always decreasing (it diverges for $v\to 0$). Nevertheless, the time dependence of their normalised moments are the same.
 \label{fig2}  }
\end{figure}

\begin{figure}[ht]
  \includegraphics[width=12.0cm]{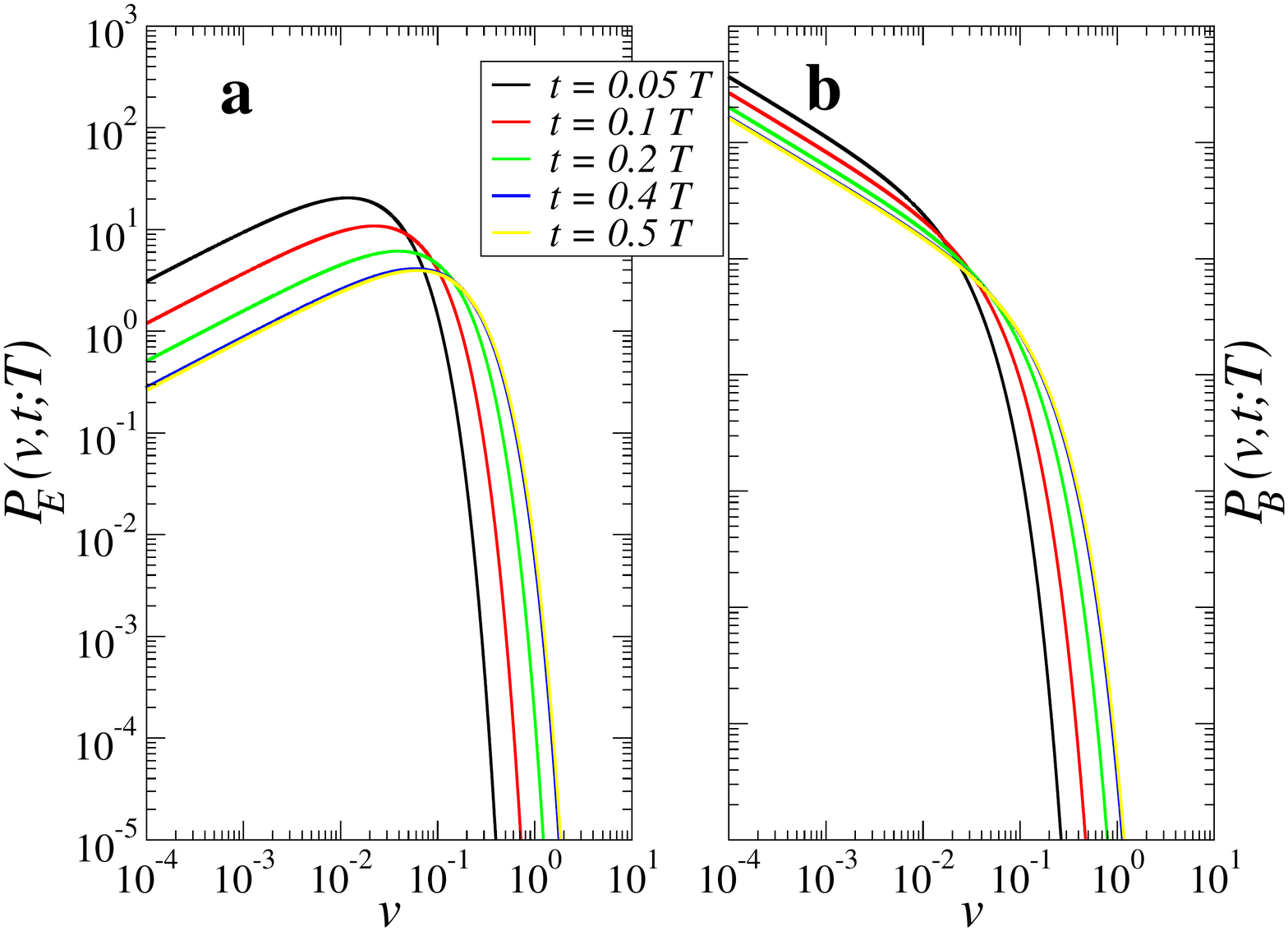}
   \caption{
  Distribution of excursion  $P_E(v,t;T)$ (a) and bridge $P_B(v,t;T)$ (b) shape, reported in Eqs.~(\ref{ExcDist}), and~(\ref{BriDist}). The distributions are computed for $\mu=-0.5$, $k=1$, and $\sigma=1$, for $T=1$ and different times $t$. Note that the distributions (both are gamma distributions) differs for the algebraic initial part. For the excursion the probability increase, from $v=0$, and attains a most probable value, while for the bridge, the distribution is always decreasing (it diverges for $v\to 0$). Nevertheless, the time dependence of their normalised moments are the same.
 \label{fig3}  }
\end{figure}

The computation of the bridge shape distribution follows exactly the same lines of what done for the excursion, with the difference that now we have to use the reflecting propagator in place of the absorbing one. That is:
\[
P_B(v,t;T) =\lim_{\epsilon\to0} \frac{P_+(v,t;\epsilon,0)P_+(\epsilon,T;v,t)}{\int_0^\infty  P_+(u,t;\epsilon,0)P_+(\epsilon,T;v,t) du}.
\]
Again, considering the small $\epsilon$ behaviour of $P_+$ and simplifying the factors in common between numerator and denominator, one gets the simplified kernel:
\[
K_2(v) = \exp\left[-\omega(t,T) v\right] v^\mu,
\] 
and the corresponding bridge distribution (see Figs.~\ref{fig2} and~\ref{fig3}, right panels)
\begin{equation}
P_B(v,t;T) = \frac{K_2(v)}{\int_0^\infty K_2(u)du} = \frac{ \omega(t,T)^{1+\mu}}{\Gamma(1+\mu)}e^{-\omega(t,T) v} v^\mu\label{BriDist}
\end{equation}
which is again a Gamma distribution, with moments:
\[
\langle v(t)^n \rangle_B = \frac{(n+\mu)(n-1+\mu)...(1+\mu)}{\omega(t,T)^n}=\frac{\Gamma(n+1+\mu)}{\Gamma(1+\mu)} \omega(t,T)^{-n}.
\]
The average bridge shape is:
\begin{equation}
\langle v(t) \rangle_B  = \frac{1+\mu}{\omega(t,T)} = \frac{(1+\mu) \sigma^2}{2 k } \frac{(e^{-k t}-1)(e^{-k (T-t)}-1)}{1 - e^{-k T}}\label{avebr}
\end{equation}
Note that, as for the two propagators $P_+$ and $P_-$, one can swap from the bridge to the excursion statistics, substituting $\mu$ with $-\mu$.

As a result, the average shape of the bridge is proportional to the shape of the excursion (the same holds for every higher order moment). This was in principle not obvious, since the bridge can be seen as  a convolution of one or more excursions, whose sum of durations is the duration of the bridge. On the other hand, it is obvious that the amplitude of the excursion is larger than the amplitude of the bridge, since the first is "forced" to stay away from the $v=0$ axis.

An other interesting observation, is that again the normalised shape of the bridge does not depend on $\mu$. Recalling that $P_+$ coincide with the solution $P_1$ of the CIR FP equation for $\mu>0$, it turns out that the bridge for the "steady-sliding" regime $c>\frac{\sigma^2}{2k}$ has the same normalised shape, irrespectively of the value of the drive $c$. However, the probability to observe a bridge of duration $T$ during the stochastic dynamics is vanishingly small (with $\epsilon$) for $\mu>0$.

\section{Constrained stochastic equations for bridge, excursion and meander: Doob's h-transform}~\label{DoobsSec}

In order to further investigate the equivalence of the normalised shape of excursion and bridge, we consider the stochastic equations for the respective constrained trajectories. To this aim, we make use of the Doob's h-transform, which allows to identify the correct drift terms for such constrained stochastic equations. For a very brief introduction of Doob's h-transform see~\cite{Majumdar2015}, as well as~\cite{Mazzolo2018} for interesting variations. In the present case, the idea behind Doob's transform, is to consider the time derivative of the bridge or the excursion probability, named here respectively $P_B$ and $P_E$. Both are in the form:

\[
	Q = {}\lim_{\epsilon\to 0} \frac{P(v,t ; \epsilon) P(\epsilon,T-t;v)}{P(\epsilon,T;\epsilon)},
\]

where $P(v,t;u)$ is the probability for the desired constrained process to assume the value $v$ at time $t$, knowing that it started at $u$ at time $0$. For the bridge, the probability $P$ is the "reflecting" propagator $P_+$, while for the excursion $P$ is the "absorbing" propagator $P_-$. In both cases, the function $P$ satisfy the FP equation associated to the ABBM/CIR stochastic process (and the corresponding adjoint, backward equation). It turns out that $Q$ also satisfies a FP equation, which reads:
\begin{equation}
\partial_t Q = -\partial_v \left[\left(k(c-v) +  \sigma^2 v \lim_{\epsilon\to 0}\partial_v \log P(\epsilon,T-t;v) \right)Q\right] + \frac 12 \partial^2_{v} \left( \sigma^2 v Q \right)  \label{constrainedFP}
\end{equation}

Eq.~(\ref{constrainedFP}) has a form similar to the original FP equation, but with an extra drift term, where the constrained propagator $P$ enters. This means that $Q$ is the propagator for a stochastic process governed by the equation:
\[
dv = \left[ k(c-v) +  \sigma^2 v \lim_{\epsilon\to 0}\partial_v \log P(\epsilon,T-t;v) \right] dt + \sigma \sqrt{v} \,dW_t.
\]
This result represents is an interesting "short-cirtuit" between the description of the stochastic process in terms of its FP equation, and the SDE describing its trajectories, two approaches that, although equivalent, usually proceeds in parallel, without intersections. 

 The explicit expression of the extra drift, which can be computed using the small $\epsilon$ expansion in Eqs.~(\ref{Rsmallx}) and~(\ref{Asmallx}), gives an intuitive interpretation of our results. For the bridge, that is for $P=P_+$, it reads:
\[
- v \left[ \frac{ 2 k  }{e^{k(T-t)}-1}\right].
\]
This term diverges when $t$ is approaching $T$ as
$-\frac{ 2 v}{T-t}$. The effect of this non stationary drift term is to force the trajectory toward the value $v=0$ at time $T$, as requested by the constraint of the bridge.

On the other hand, for the excursion, the extra drift term has to be computed using the "absorbing" solution $P=P_-$, and the corresponding small expansion Eq.~(\ref{Asmallx}). The result is:
\[
- v\left[ \frac{ 2 k }{e^{k(T-t)}-1}  \right] -\sigma^2 \mu
\]
Again we find the term imposing the passage of the trajectory at $v=0$ for time $T$, but now we have a constant repulsive drift term that keep the trajectory away from $v=0$ at all time $t$. This last term is the effect of the excursion constraint to not touch the $v=0$ axis.

The two stochastic equations can be written as:

\begin{equation}
dv =  \left\{ A_{B,E} -   v\,k \coth \left[\frac{k (T-t)}2\right] \right\} dt   + \sigma \sqrt{v} \,dW_t \;\;\text{(bridge, excursion),}\label{Doobsde}
\end{equation}

where $A_B = kc  = (1+\mu)\sigma^2/2$ for the bridge and $A_E = \sigma^2 -kc  = (1-\mu)\sigma^2/2$ for the excursion. The relation between bridge and excursion can be fully appreciated considering the rescaled variable $u=\frac{v}{A_{B,E}}$ one gets:
\begin{equation}
du =  \left\{ 1 -   u\,k \coth \left[\frac{k (T-t)}2\right] \right\} dt  +\frac{\sqrt{2\,u}}{\sqrt{1 \pm \mu}} \,dW_t \;\;\text{($+$ bridge, $-$ excursion),}\label{Doobsdemu}
\end{equation}

{Recalling that the equation for the bridge is valid even for positive values of $\mu$, we observe, from this formulas, that the excursion of parameter $\mu = \mu_0 <0$ is identical to a bridge of parameter $\mu = -\mu_0 >0$.} In terms of dimensionality $\delta = 2(\mu+1)$ of the underlying OU process, this means that an excursion (of its squared modulus) for a value of $\delta<2$ is identical to a bridge (of the squared modulus) of a OU of dimension $\delta'=4-\delta$. For instance, the excursion for the squared modulus of a one dimensional OU process is identical to the bridge for the squared modulus of a three dimensional OU process, which, in turn, is the sum of three independent one dimensional bridges:
\begin{equation}
Y_E(t) = Y_{B,1}(t)+Y_{B,2}(t)+Y_{B,3}(t) = Y^{\delta=3}_B
\end{equation}
Here $Y_E$ and $Y_B$ are, respectively, the excursion and the bridge, for the process $Y=y^2$, where $y$ is a one dimensional OU process, while $Y^{\delta=3}_B$ is the bridge of the squared modulus for a $\delta=3$ OU process.
This is the generalization of the known equivalence between the excursion of the Wiener process and a three dimensional Bessel process~\cite{Pitman2018}:
\begin{equation}
X_E(t) = \sqrt{[X_{B,1}(t)]^2+[X_{B,2}(t)]^2+[X_{B,3}(t)]^2} = Z^{\delta=3}_B,\label{BroExc2Br}
\end{equation}
where here $X_E$ and $X_B$ are, respectively, the excursion and the bridge of a Brownian process in $\delta=1$, while $Z^{\delta=3}_B$ is the bridge of a $\delta=3$ Bessel process.

We also remark, from Eq.~(\ref{Doobsdemu}), that although, as shown before, every moment of the the excursion is proportional to the corresponding moment of the bridge, the two processes are not proportional, i.e. one can not obtain one from the other with a simple rescaling of the stochastic variable. In fact, as can be observed from the explicit expressions of the moments, the proportionality constant between moments depends on the moment order.

Coming back to the full parametrized Eq.~(\ref{Doobsde}), note that, in the limit of $T\to \infty$ the stochastic equation for the bridge recovers the original stochastic ABBM/CIR equation, as it should, since $\lim_{x\to\infty} \coth(x) = 1$. The equation, written using the parameters $\delta,k,\sigma$ instead of $c,k,\sigma$, reads:
\begin{equation}
dv =  \left\{ \frac{\delta \sigma^2}4 - k v\right\} dt   +  \sigma \sqrt{v} \,dW_t\;\;\text{(free)}\label{GBESQ}
\end{equation}
On the other hand,  performing the same $T\to\infty$ limit for the excursion, the constant drift remains, and we get an equation, which, written using the parameters $\mu,k,\sigma$ instead of $c,k,\sigma$, reads
\[
dv =  \left\{ \frac{\sigma^2}2 (1-\mu)- k v \right\} dt   + \sigma \sqrt{v} \,dW_t \;\;\text{(infinite meander).}
\]
This is the stochastic equation for the process only constrained to never touch the axis $v=0$, i.e. is an avalanche of infinite duration. In stochastic theory this process can be considered a {\em meander} of infinite duration (the meander being a process constrained to never touch $0$ up to time $T$, regardless of its value at $T$). Note from its stochastic equation that it is exactly the original process, with the index $\mu$ changed of sign. In this sense, we can say that the a "steady sliding" trajectory of the ABBM-CIR model (that is a solution of the sde equation with $\mu>0$) is nothing but the equation for an an infinite avalanche (or infinite meander) in the mirrored "stick-slip" regime, that is a solution of the sde with opposite (negative) value of $\mu$. 

Coming back to the  bridge and excursion shape, we can consider the average of  Eq.~(\ref{Doobsde}), luckily obtaining a closed equation:

\[
d\langle v\rangle = \left\{ (1\pm\mu)\frac{\sigma^2}2 -  \langle v\rangle \,k \coth \left[\frac{k (T-t)}2\right] \right\} dt 
\]
where we dropped the noise term, since we are following It\^o's scheme. Solving such equation for $\langle v(0)\rangle = 0$ gives again Eqs.~(\ref{CIRavex}) and~(\ref{avebr}).

\subsection{Radial Ornstein-Uhlenbeck and Bessel process}

As explained before, the square root of the ABBM/CIR process is a Generalised Bessel process (GBES), which can describe the squared modulus of $\delta$-dimensional OU process. In this case, the square root pof the process is named radial Ornstein-Uhlenbeck (ROU). More precisely, if we consider $\rho=\sqrt{v}$ and apply It\^o's lemma to Eq.~(\ref{Doobsde}) we get
\begin{equation}
d\rho = \left[\frac{\sigma^2 (1\pm 2\mu)}{8\rho} - \frac 12\, \rho\, k \coth \left( \frac{k(T-t)}2 \right) \right] dt+\frac\sigma2 dW_t,\label{BesselBriExcSde}
\end{equation}
which, depending on the choice of the sign before $\mu$, is the sde for the Bridge ($+$), Excursion ($-$), Infinte meander ($\lim_{T\to\infty} (-)$) of a radial Ornstein-Uhlenbeck (ROU) process in dimension $\delta=2(\mu+1)$, whose free equation is $\lim_{T\to\infty} (+)$:
\begin{equation}
d\rho = \left[\frac{\sigma^2 (\delta-1)}{8\rho} - \frac 12\, \rho\, k  \right] dt+\frac\sigma2 dW_t,\label{ROU}
\end{equation}

In the limit of $k\to 0$, Eq.~(\ref{BesselBriExcSde}) recovers the analogous Bridge/Excursion sde for the standard Bessel process
\[
dw = \left[ \frac{(1\pm 2\mu)}{2\, w} - \frac w{2(T-t)} \right] dt+ dW_t,
\]
where we changed the time $\sigma^2 t/4\to t$ (and $\sigma dW_t/2\to dW_t$) in order to get for the free ($\lim_{T\to\infty} (+)$) sde the standard form:

\begin{equation}
dw = \frac{\delta-1}2 \frac{dt}{w}+ dW_t.\label{BESstandard}
\end{equation}

Incidentally, note that this equation shows that care has to be used for the case $\delta=1$ in identifying the Bessel process with the sde for the absolute value of the Brownian process. One can understand this point, considering the formal application of It\^o lemma from the OU sde $dx = -k\,x\, dt + \sigma dW_t$ to the function $\rho=f(x)=|x|$, where one has to use $f'(x)=\sign(x)$ and $f''(x)=2 \delta(x)$. This gives:
\[
d\rho = \left[-k\, x \sign(x) + \delta(x) \sigma^2\right] dt + \sigma \sign(x) dW_t. 
\]
Now, since $x\sign(x) = |x|$, and since $\sign(x)dW_t$ is a new Wiener process (the sign of $W_t$ is uncorrelated with the sign of $x_t$, since we are in It\^o stochastic framework) this can be written in closed form as
\[
d\rho = -k\,\rho\,dt + \sigma \,dW_t + dL_t,
\]
where $L_t$ is the {\em local time of the process in $0$}~\cite{Bjork2019} defined as
\[
L_t = \int_0^t \delta(w_s)\,\sigma^2\,ds.
\]
Apart for the $\sigma^2$ factor, the local time is the function measuring the time spent by the process at the value $0$. This result is exact and, in the general case, it is known as the It\^o-Tanaka formula. 
(See \cite{Grebenkov2020} for a recent application of local times in physics).

The shape of the radial OU (ROU) bridge (and of the Bessel bridge, in the limit $k\to 0$)  is identical to that of the corresponding excursion, as happens for the ABBM/CIR case. However, the computation can not be performed easily from the sde Eq.~(\ref{BesselBriExcSde}), since averageing both side of the equation does not result in a closed equation for the first moment, due to the non linearity of the equation.

In order to compute the average shapes (and their moments) we have to use the explicit expression through  the solutions of the associated FP equation. Using the solutions of the ABBM/CIR equation, we get that:

\[
P_{ROU}(\rho,t;\rho_0) = \frac{1}{2\rho} P_{CIR}(\rho^2,2(1\pm\mu) t;\rho_0^2). 
\]

The computation, then, follows exactly the same lines as before, with the reflecting 
\[
P_{ROU,+}(\rho,t;\rho_0) =\frac{1}{2\rho} \lambda \exp\left[-\lambda\cdot(\rho^2+\rho^2_0 e^{-k t})\right] \cdot \left(\frac{\rho}{\rho_0 e^{-k t/2}}\right)^{\mu} \cdot I_{\mu}\left( 2\lambda \rho\,\rho_0 e^{-k t/2} \right)
\]
and the absorbing solutions

\[
P_{ROU,-}(\rho,t;\rho_0) =\frac{1}{2\rho} \lambda \exp\left[-\lambda\cdot(\rho^2+\rho^2_0 e^{-k t})\right] \cdot \left(\frac{\rho}{\rho_0 e^{-k t/2}}\right)^{-\mu} \cdot I_{-\mu}\left( 2\lambda \rho\,\rho_0 e^{-k t/2} \right)
\]
The result for the distribution of the ROU excursion $Q_E$ and the bridge $Q_B$:
\begin{eqnarray}
Q_E(\rho,t,T) &=& \frac{2\,\omega(t,T)^{1-\mu}}{\Gamma(1-\mu)} \exp\left[-\omega(t,T)\rho^2\right]\rho^{-2\mu+1}\\
Q_B(\rho,t,T) &=& \frac{2\,\omega(t,T)^{1+\mu}}{\Gamma(1+\mu)} \exp\left[-\omega(t,T)\rho^2\right]\rho^{2\mu+1},
\end{eqnarray}
where $\omega$ has the same form of the ABBM/CIR case, Eq.~(\ref{omega}).
The corresponding moments are:
\begin{eqnarray}
\langle \rho(t)^n \rangle_E &=& \frac{\Gamma(1-\mu+n/2)}{\Gamma(1-\mu)} \omega(t,T)^{-n/2}\\
\langle \rho(t)^n \rangle_B &=& \frac{\Gamma(1+\mu+n/2)}{\Gamma(1+\mu)} \omega(t,T)^{-n/2}.
\end{eqnarray}

Again, the expression are the same, except for their amplitudes, as for the CIR process.

Moreover, in the limit $k\to 0$, these expressions give the moments for the Bessel excursion and bridge shapes, where $\omega$ is in that case
\begin{equation}
\omega(t,T)=\frac{2}{\sigma^2} \frac{T }{(T-t)\,t},
\end{equation}
which again benefit of the same universality, when in their normalised form.

\section{Conclusions} \label{ConclusionsSec}

In this paper, we collect some known exact results for the Cox-Ingersoll-Ross (CIR) model, which can be relevant for the physics of Barkhausen noise, as modelled by the Alessandro-Beatrice-Bertotti-Montorsi (ABBM) model. In particular, the procedure of {\em time change}, which connect the ABBM model with the CIR process, is briefly described. The CIR process, in turn, is equivalent to the a generalised Squared Bessel process,  (GBESQ), a stationary, mean reverting generalization of the Squared Bessel process (BESQ). In fact, the Bessel (BES) process describes the modulus of a $\delta$ dimensional Brownian process (BRO), while the Generalized Bessel (GBES) process represents the modulus of an Ornstein-Uhlenbeck (OU) process in $\delta$ dimensions, and in this case is also named Radial Ornstein-Uhlenbeck (ROU) process. The connections between all the aforementioned stochastic processes are summarized in Fig.~\ref{FigStochasticProcesses}.

Leveraging such results, we adressed the exact computation of the average avalanche shape, which in stochastic theory is the average shape of an excursion of the process. In order to compute this quantity, we used the absorbing solution of the CIR Fokker-Planck equation, computed by Feller in 1951~\cite{Feller1951}, and we obtained the exact distribution of the avalanche shape, for every values of the model parameters (previous results being restricted to specific, limiting values). Extending the computation to the case of bridges, that is the generic sequence of avalanches whose total duration sum to a fixed time $T$ (named here multi-avalanche), we obtained an interesting similarity. Although the bridge shape distribution differs from the excursion case, the time dependence of their moment is identical: normalizing moments of the two shapes by an arbitrary measure of their amplitude (e.g. their average area or the average maximum value), one obtains the same exact expression as a function of time.

 \begin{figure}[ht]
   \includegraphics[width=12.0cm]{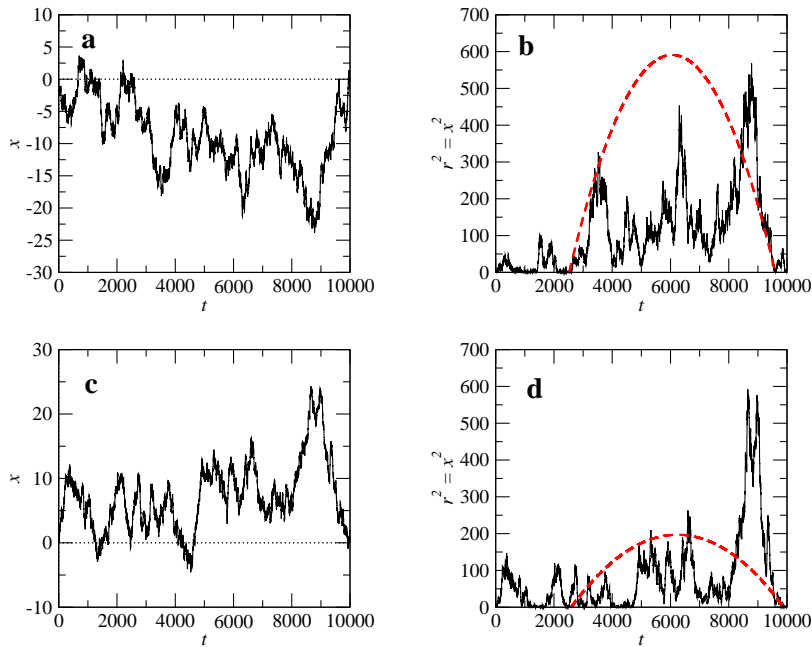}
   \caption{Representation of two trajectories for a one dimensional Brownian motion (panels a and b) and the corresponding squared distance as a function of time (panels b and d): an excursion (panel b) and a bridge (panel d) of similar durations are compared with their respective average shape (red dashed line).\label{figBrownian}  }
\end{figure}

Considering the connection of the Bessel process with Brownian process, the result can be recast in terms of the average shape of the trajectory for the distance of a Brownian particle from its starting position (the origin). In the pictorial representation of Fig.~\ref{figBrownian}, two Brownian processes are shown (panels a and c). The distance from the origin of the two trajectories, is also reported (in panels b and d), where one can identify an excursion (panel b) and a bridge (panel d) of similar duration. As proven in this paper, the average shape of these two quantities differ for a multiplicative constant only: the normalised shape of the average distance between two returns does not depend on the number of intermediate passages to the origin. 

This result appears quite counter intuitive: one could expect that average shapes of trajectories subjected to  a different constraint (touching or not touching the zero value), should differ, in general. Excursion trajectory should fly farther from the zero-axis than bridges, which can wander freely nearby zero even in the middle of the bridge duration. A refined intuition could take into account the algebraic decay of the avalanche durations in the "scale-free regime", that is for not too large avalanches. In this case, one could suspect that the multi-avalanche is in fact dominated by a single large avalanche, whose duration is about the whole multi-avalanche duration, making avalanches and multi-avalanches essentially the same stochastic object, apart from some small avalanches "decorating" the typical trajectory near the border (i.e. at the beginning and at the end of the bridge). Nevertheless, the equivalence between excursion and bridge shapes holds for every value of durations, hence in every regime of the duration distribution. In other words, since the duration distribution of avalanches decay exponentially at large durations ($T\gg 1/k$), we know that the typical multi-avalanche trajectory for such large total durations is made by several avalanches of comparable durations. This is exactly the opposite scenario that the naive hand-waving explanation invokes in order to explain the equivalence of avalanche and bridge shape.

To our knowledge, in the mathematical literature, shapes of bridge or excursion has not been considered explicitly. Analogies between bridge and excursion of Brownian process has been proved, notably regarding the distribution of the range (maximum minus minimum values) or an equivalence between the excursion and a bridge, shifted so to start from its minimal value~\cite{Vervaat1978} (Brownian process is a symmetric process, so the minimum of the bridge can be negative). Nevertheless, for the CIR process, we find that, despite the shape distributions of bridge and excursion differ, the normalised moments of the shapes share the same dependence with time. This specific result, applies to all the stochastic processes in the Generalised Bessel family considered here.

The generality of this phenomenon, beyond the diffusion processes considered here (which encompass multidimensional Brownian (and OU) process) is not clear. One can wonder if other diffusion processes share this equivalence. In general, the shape distributions of bridges and excursions depends on the reflective and absorbing propagators, which are different. Nevertheless, being the solutions of the same FP equation, the two propagators are related. In an interesting paper~\cite{Grosche1993}, it has recently be noted that a quite simple relation exists between the Laplace transforms of propagators subjected to different boundary conditions:

\begin{equation}
\hat P_{a}(x,x_0;\lambda) = \hat P(x,x_0;\lambda)-\frac{\hat P(a,x_0;\lambda) \hat P(x,a;\lambda)}{\hat P(a,a;\lambda)},~\label{ref2abs}
\end{equation}
where $P$ is the free propagator, while $P_{a}$ is the solution of the FP where an absorbing boundary condition has been introduced at $x=a$, and the hat symbol stand for the usual time Laplace transform
$\hat f(t) = \int_0^\infty \exp(-\lambda t)f(t)dt$. However, since the observed equivalence resides in the temporal dependance of the distributions, the relation Eq.~(\ref{ref2abs}) between Laplace transforms is not easy. 

On the other hand, one can try to attack the problem via the Doob's transformed sde for the bridge and the excursion. They usually differ for a drift term, the repulsive force which keep the excursion far from the absorbing boundary. The equivalence observed in this paper indicates that this repulsive drift term does not change the temporal behaviour of the average (and moments) of the processes, rather it only affects its amplitude. In~\cite{Pinsky1985} (see~\cite{Mazzolo2018} for a less formal derivation), it is shown that, for certain diffusion processes, the drift term associated to a Doob's tranform involves the first eigenfunction of the original Fokker-Planck operator. Further investigations are needed in order to understand if these result could be useful to generalise the observed equivalence between bridge and excursion shape in ABBM/CIR/Bessel processes to other diffusion processes.

Recently, a generalization of Doobs's transform has been introduced in order to investigate the equivalence of ensembles in non-equilibrium statistical physics~\cite{Chetrite2013,Chetrite2015}. In that case, the set  of trajectories subject to a constrain, forcing the system to very atypical paths, plays the role of the micro-canonical ensemble (where only configurations with the same energy are considered). On the other hand, one can consider a "canonical ensemble", where the constrained is removed, but all the trajectory are weighted with a penalty depending on the constraint and a free parameter (as in the canonical ensemble, where every configurations contribute to the ensemble, but with a Gibbs weight, which depends on the energy and the temperature).  In the limit of large times, and under several conditions, authors of~\cite{Chetrite2015} proof a logarithmic, asymptotic identities between the distributions of the two ensembles. The analogy with equilibrium statistical physics is deeply grounded on the theory of large deviations~\cite{Touchette2009}, which represents the mathematical framework where both theories can be successfully cast. 

In the present case, however, for a much more restricted class of diffusion processes, we observed a correspondence between a special class of paths (excursion) and an unrestricted one (bridge): however the two ensemble do not have the same distribution, but they rather share an identity in the time evolution of their moments. Such identity, however, applies at finite time, without the need of an asymptotic limit. Nevertheless, it would be interesting to further investigate our problem in terms of the work~\cite{Chetrite2015}, for instance considering a "canonical ensemble" of bridges, where the trajectories are weighted according to the number of returns to zero, and then try to relate the resulting distribution with that of excursions. Such a research program could maybe shed light on the limitations or generality of our result.  

However, note that one of the conditions required for the ensemble equivalence is a "non-condensation" property of the dynamics. In~\cite{Szavits-Nossan2015} such a condition is investigated on  some diffusion processes (BRO, OU and CIR) considered in the present work. The {non-condensation} property deals with the effect of the constraint considered (the final generic final value at time T) on the whole temporal evolution, rather than concentrated at the temporal extreme of the evolution, where the constraint has to be verified. In the case considered there, "non-condensation" phenomenon is related to the temporal form of the non-stationary Doob drift term of the bridge. While for BRO, the average bridge from $0$ to a value $a>0$ is a straight lines connecting the two points, for OU and CIR, the average constrained trajectory reaches the target exponentially fast near the arrival, while is close to the unconstrained average trajectory for previous times. Note that this behaviour is not surprising, considering the non linear time change relation between BRO and OU, as well as the relation between CIR and the squared modulus of $\delta$-dimensional OU. Nevertheless, the different time parametrization change completely the asymptotic distribution of the processes, since in this case for BRO the ensemble equivalence holds, while for OU and CIR do not. It would be interesting to investigate the effect of time change transformations in the more general framework of~\cite{Chetrite2015}, in order to shed light on the minimal conditions needed for the equivalence to apply, as well as to identify classes of processes sharing the same equivalence properties. 

Finally, a more practical and preliminary research direction could be to take in account simple numerical investigations on specific stochastic processes. {For instance, one can try to consider the effect of time correlation, beyond the simple (one-dimensional) Markov assumption. This can be obtained with colored noise, or with the introduction of other inertial or memory effects. The average avalanche shape (excursion) has been considered for a random accelerated particle~\cite{Colaiori2004,Baldassarri2003} or for a more sofisticated memory kernel~\cite{Zapperi2005}. In both cases, asymmetric avalanche shapes have been observed (see also~\cite{Baldassarri2019} for other asymmetric avalanche shapes in granular friction experiments), but the shape of multiavalanches have not been studied in all these cases.}

If the equivalence between excursions  and bridges is believed to hold, it can be exploited in order to measure avalanche shapes with very large durations (which requires huge time series) considering the much more frequent corresponding bridges. For shorter durations, instead, checking the equivalence between bridge and excursion can eventually represent a refined check in order to assess the limits of the chosen stochastic modellization for the fluctuating physical process in study.

\section*{Acknowledgments}
I am grateful to M. Gianfelice for useful discussions and the careful reading of a preliminary version of the manuscript.

\appendix

\section{Connections between ABBM/CIR and other stochastic processes}

\label{connections}

\subsection{Squared Bessel Process}
First consider Eq.~(\ref{CIRsde}) 
in the limit $k\to 0$ with $\delta \equiv \frac{4kc}{\sigma^2}$ fixed, the CIR sde becomes:
\be
dv = \frac{\delta \sigma^2}4 dt + \sigma \sqrt{v} dW_t.\label{CIRBESQ}
\ee
After a simple rescaling $X \equiv 4v/\sigma^2 $, we recover the Squared Bessel  (BESQ) process:
\begin{equation}
dX = \delta dt + 2 \sqrt{X} dW_t.\label{BESQstd}
\end{equation}
The Squared Bessel process is the square of a Bessel process, which in turn, for integer $\delta$, describes the modulus of the $\delta$-dimensional Wiener process.  In fact, given $\delta$ dimensional Wiener process of independent coordinates $W_i(t)$, consider the (rescaled) squared modulus { $v = \frac{\sigma^2}4 \sum W_i(t)^2$,} then applying It\^o's lemma, one gets
\begin{equation}
dv = \frac{\delta\sigma^2}4\,dt +\frac{\sigma^2}2 \sum_i W_i(t) dW_i(t). \label{preSqBessel}
\end{equation}
Since it can be proved that
$$
\sum_i W_i(t) dW_i(t) \stackrel{d}= \sqrt{\sum_i W_i(t)^2}\, dW_t,
$$
where $W_t$ is an independent scalar Wiener process, the noise term in Eq.~(\ref{preSqBessel}) is readily identified with  $\sigma \sqrt{v} dW_t$, recovering Eq.~\eqref{CIRBESQ}.

As we will see soon, there exists a more general connection with Bessel processes also for $k\neq 0$.

\subsection{Rayleigh process}
It is possible, via the so called Lamperti transform, to get rid of any multiplicative noise term. Here we save the parameter $\sigma$ and we consider 
$$
du = \frac{dv}{2\sqrt{v}},
$$
that is we consider the new variable $u=\sqrt{v}$. Using It\^o's lemma, it is easy to write the stochastic equation for $u$:
\begin{equation}
du =\frac 12\left[\frac{4kc-\sigma^2}{4 u}- k  u\right]dt+\frac{\sigma}2\,dW_t \label{rayleigh}
\end{equation}
This process is a special case of the so called Rayleigh process:
$$
dZ = \left(\frac AZ+B\,Z\right)dt+\sigma dW_t
$$
where $Z=2u$, $B=k/2$ and $A=4kc-\sigma^2$.

\subsection{Ornstein-Uhlenbeck process and Generalized Squared Bessel Process}

For $4kc-\sigma^2 = 0$, i.e. $c=\frac{\sigma^2}{4k}$, the Eq.~(\ref{rayleigh}) for $u$ becomes an Ornstein-Uhlenbeck process:
$$
du = - \frac 12 k \, u \,dt+\frac{\sigma}2\,dW_t
$$
Note that now the the stochastic variable $u$ can be negative, but this is not a contradiction with the original positive CIR variable, since it is $v=u^2$.

Nevertheless, there is a more general relation between CIR process and OU process. Let's consider a $\delta$-dimensional OU-process:

$$
dY_i = -\frac k2 Y_i dt + \frac{\sigma}2 dW_i(t)
$$
where, as before, $i=1,...,\delta$ and $W_i(t)$ are $\delta$ independent Wiener processes.

Again, considering $v = \sum_i Y_i^2$ and applying It\^o's lemma one has:
$$
dv = k\left(\frac{\delta\sigma^2}{4 k}-v\right)\,dt +\sigma\sum_i Y_i dW_i(t),
$$
whose noise term is equivalent to 
$\sigma \sqrt{v} dW_t$, because, as before,  $\sum_i Y_i dW_i \stackrel{d}= \sqrt{\sum_i Y_i^2}\, dW_t$. The result it that 
we can identify the CIR process with the stochastic equation for the squared modulus of a $\delta$ dimensional OU-process, where and $c=\frac{\delta\sigma^2}{4k}$. 

Note that it is possible to recover this result, that is the equivalence of the generic ($k\neq 0$) CIR model with the Generalize squared Bessel process, starting from the case $k=0$ considered above to get the squared Bessel process, and then performing the same deterministic time change that has been shown to lead from the Wiener process to the Ornstein-Uhlenbeck process.

Summarizing, as the Bessel process describes the modulus of a $\delta$-dimensional Brownian process, so the CIR process describes the squared modulus of a $\delta$-dimensional OU-process, which is also called Generalized Squared Bessel Process (GBESQ). In Fig.\ref{FigStochasticProcesses} we summarize all the stochastic processes mentioned in the paper and their relations.

\section{Generating function for moments of the CIR process} \label{generating}

A very interesting result is the explicit computation of the generating functions for the moments of the velocity as well as the moments of the displacement $x(t)=\int_0^t v(u) du$ from the (reflecting) propagator of the CIR model (the proof of the following go beyond the scope of this paper). It can be shown that (see~\cite{Lamberton1996}, p.130, or~\cite{Jeanblanc2009}, p.361):

\be
\langle \exp(-\alpha v(t) -\beta x(t))\rangle_{v_0} = \exp\left\{-A(\alpha,\beta;t)-v_0 B(\alpha,\beta;t)\right\},
\ee
where the subscript $v_0$ explicitly indicates the dependence of the average from the initial condition $\langle \cdot \rangle_{v_0}$, and
\begin{eqnarray}
A(\alpha,\beta;t,v_0) &=& -\frac{2kc}{\sigma^2}\ln\left[\frac{2\gamma e^{(\gamma+k)t/2}}{\sigma^2 \alpha(e^{\gamma t} -1) + \gamma(e^{\gamma t}+1)+k(e^{\gamma t}-1)}\right] \\
B(\alpha,\beta;t)&=&\frac{\alpha\left[\gamma+k+e^{\gamma t}(\gamma-k)\right]+2\beta(e^{\gamma t}-1)}{\sigma^2\alpha(e^{\gamma t}-1)+\gamma(e^{\gamma t}+1)+k(e^{\gamma t}-1)}
\end{eqnarray}

where
\[
\gamma=\sqrt{k^2+2\sigma^2\beta}.
\]
From this expressions, the explicit expression for the generic moment can be computed with a simple differentiation:
\begin{equation}
\langle v^m x^n\rangle_{v_0} =(-1)^{n+m}\left. \frac{\partial^{m+n}}{\partial \alpha^m \partial \beta^m} \exp\left\{-A(\alpha,\beta;t)-v_0 B(\alpha,\beta;t)\right\}\right|_{\alpha=0,\beta=0}
\end{equation}

In particular, for $m=0$, the moments for the velocity are

\begin{equation}
\langle v^n\rangle_{v_0} = (-1)^n \frac{\partial^n}{\partial \alpha^n}\left. \left\{ \left( \frac {\lambda(t)}{\alpha+\lambda(t)}\right)^{\mu} \exp\left(-\frac 12 \frac{\alpha  v_0 \,\lambda(t)e^{-k t}}{\alpha + \lambda(t)} \right)\right\}\right|_{\alpha=0},
\end{equation}
where as before $\mu=2kc/\sigma^2$ and $\lambda(t) = 2k/[\sigma^2(1-e^{-kt})]$.
For $n=0$, the moments for the displacement are
\begin{equation}
\langle x^m\rangle_{v_0} = (-1)^m \frac{\partial^m}{\partial \beta^m} \left\{e^{k^2ct/\sigma^2}\left.\left(\cosh\frac{\gamma t}2 + \frac k\gamma\sinh \frac{\gamma t}2\right)^{-2kc/\sigma^2}\exp\left(\frac{-2\beta v_0}{k+\gamma \coth \frac{\gamma t}2}\right)\right\}\right|_{\beta = 0},
\end{equation}
where $\gamma^2=k^2+2\beta\sigma^2$

\bibliography{bridge-excursion}

\end{document}